\documentclass{emulateapj}

\slugcomment{accepted for publication in ApJ}

\shorttitle{Vega Debris Disk}
\shortauthors{Su et al.}

\newcommand{\um}{${\rm \mu m}$~}
\newcommand{\mm}{${\rm \mu m}$}

\begin{document}

\title{The Vega Debris Disk -- A Surprise from {\it Spitzer}\altaffilmark{1}}
\author{K. Y. L. Su\altaffilmark{2}, 
  G. H. Rieke\altaffilmark{2}, 
  K. A. Misselt\altaffilmark{2},
  J. A. Stansberry\altaffilmark{2}, 
  A. Moro-Martin\altaffilmark{2,4}, 
  K. R. Stapelfeldt\altaffilmark{3}, 
  M. W. Werner\altaffilmark{3},
  D. E. Trilling\altaffilmark{2},  
  G. J. Bendo\altaffilmark{2}, 
  K.  D. Gordon\altaffilmark{2},   
  D. C. Hines\altaffilmark{5}
 M. C. Wyatt\altaffilmark{6},   
  W. S. Holland\altaffilmark{6},
  M. Marengo\altaffilmark{7},
  S. T. Megeath\altaffilmark{7},
  G. G. Fazio\altaffilmark{7} 
  }   

\altaffiltext{1}{Based on observations with NASA Spitzer Space
  Telescope, which is operated by the Jet Propulsion Laboratory, 
  California Institute of Technology under NASA contract 1407.} 
\altaffiltext{2}{Steward Observatory, University of Arizona, 933 N
  Cherry Ave., Tucson, AZ 85721; ksu@as.arizona.edu}
\altaffiltext{3}{JPL/Caltech, 4800 Oak Grove Drive, Pasadena, CA 91109}
\altaffiltext{4}{Max-Planck-Institut f\"{u}r Astronomie,
  K\"{o}nigstuhl 17, D-69117 Heidelberg, Germany}
\altaffiltext{5}{Space Science Institute, 4700 Walnut St. Suit 205, Boulder,
  Colorado 80301} 
\altaffiltext{6}{UK Astronomy Technology Centre, Royal Observatory,
  Blackford Hill, Edinburgh EH9 3HJ} 
\altaffiltext{7}{Harvard-Smithsonian Center for Astrophysics, 60
  Garden Street, Cambridge, MA 02138}

\begin{abstract}

We present high spatial resolution mid- and far-infrared images of the
Vega debris disk obtained with the Multiband Imaging Photometer for
{\it Spitzer} (MIPS). The disk is well resolved and its angular size
is much larger than found previously. The radius of the disk is at
least 43\arcsec~(330 AU), 70\arcsec~(543 AU), and 105\arcsec~(815 AU)
in extent at 24, 70 and 160 \mm, respectively.  The disk images are
circular, smooth and without clumpiness at all three wavelengths. The
radial surface brightness profiles follow radial power laws of
$r^{-3}$ or $r^{-4}$, and imply an inner boundary at a radius of
11\arcsec$\pm$2\arcsec~(86 AU).  Assuming an amalgam of amorphous
silicate and carbonaceous grains, the disk can be modeled as an
axially symmetric and geometrically thin disk, viewed face-on, with
the surface particle number density following an inverse radial power
law. The disk radiometric properties are consistent with a range of
models using grains of sizes $\sim$1 to $\sim$50 \mm. The exact
minimum and maximum grain size limits depend on the adopted grain
composition. However, all these models require a $r^{-1}$ surface
number density profile and a total mass of 3$\pm$1.5$\times$10$^{-3}$
$M_{\earth}$ in grains. We find that a ring, containing grains larger
than 180 \um and at radii of 86-200 AU from the star, can reproduce
the observed 850 \um flux, while its emission does not violate the
observed MIPS profiles. This ring could be associated with a
population of larger asteroidal bodies analogous to our own Kuiper
Belt. Cascades of collisions starting with encounters among these
large bodies in the ring produce the small debris that is blown
outward by radiation pressure to much larger distances where we detect
its thermal emission. The relatively short lifetime ($<1000$ years) of
these small grains and the observed total mass,
$\sim$3$\times$10$^{-3}$ $M_{\earth}$, set a lower limit on the dust
production rate, $\sim$10$^{15}$ g/s.  This rate would require a very
massive asteroidal reservoir for the dust to be produced in a steady
state throughout Vega's life. Instead, we suggest that the disk we
imaged is ephemeral and that we are witnessing the aftermath of a
large and relatively recent collisional event, and subsequent
collisional cascade.

\end{abstract}


\section{Introduction}

Vega ($\alpha$ Lyrae = HD 172167 = HR 7001, A0 V, d=7.76 pc) has been
used both as a fundamental photometric standard and as a template for
modeling of stellar atmospheres.  One of the highlights of the 
{\it IRAS} mission was 
the discovery of a large infrared excess around Vega,
which was observed originally as a photometric standard. The far
infrared emission was attributed to thermal dust emission from a disk
of debris orbiting the star \citep{aumann84}. 
Subsequent {\it IRAS} observations of other main sequence
stars in the solar neighborhood found that nearly 15\% of them exhibit
infrared excesses \citep{backman93,mannings98}. 

The dust found around these main sequence stars cannot be primordial
material left over from the star forming stage because the time scale
to remove primordial material ($\lesssim$10 Myr) is short compared to
the lifetimes of these stars (for example, Vega is estimated as 350
Myr old, \citealt{barrado98,song00}). Therefore, the dust must be
re-supplied. Second-generation dust in such systems is thought to
arise primarily from collisions between planetesimals and from
cometary activity. Ground- and space-based observations of the nearest
subsample of Vega-like stars from optical to submillimeter wavelengths
have revealed disk-like structures --- debris disks --- with density
gaps and enhancements
\citep{holland98,greaves98,koerner01,wilner02,weinberger02,clampin03}.
Debris disks are the most visible signposts of other planetary
systems, representing indirect evidence of planetary system
formation. We can learn about the diversity of planetary systems from
the study of debris disk structures. Vega is one of the closer and
brighter debris disks, which provides us an opportunity to study it in
detail spatially, as a foundation for understanding other debris
disks.

\citet{harvey84} observed Vega at 47 and 95 \um with the Kuiper
Airborne Observatory and inferred that the source of infrared emission may
be as large as 23\arcsec~in radius.
\citet{bliek94} re-analyzed the {\it IRAS} data and found that the {\it IRAS} 60
\um emission from Vega is 18\arcsec$\pm$3\arcsec~in radius. They
also suggested that the emission at 60 \um is caused by small dust
grains with size between 0.1 and 10 \mm. {\it Infrared Space
  Observatory (ISO)} observations show a
smooth, resolved, face-on disk with a radius of 22\arcsec
$\pm$2\arcsec~at 60 \mm, and 36\arcsec $\pm$3\arcsec~at 90 \um
\citep{Heinrichsen98}.
\citet{mauron98} attempted to detect Vega's disk through optical
scattered light by using linear photopolarimetry. Their upper limit
implies that the debris around Vega contains a very small number of
0.01-0.3 \um grains.
\citet{zuckerman93} detected emission by dust from Vega at 800 \um and 
suggested the presence of 100-300 \um grains to account for the submillimeter
observations.

The 850 \um map of Vega obtained by \citet{holland98} with the
Submillimeter Common-User Bolometer Array (SCUBA) on the James Clerk
Maxwell Telescope shows an extended, roughly circular structure with
dimensions of 24\arcsec$\times$21\arcsec($\pm$2\arcsec). The 850 \um
resolved emission is not uniformly distributed; instead, there is an
elongated bright central region oriented northeast-southwest with a
bright peak offset $\sim$9\arcsec~(70 AU) to the northeast of the
star's position. Observations at 1.3 mm obtained by \citet{koerner01}
and by \citet{wilner02} resolved dust emission peaks offset from the
star by 8\arcsec~to 14\arcsec~that appear to be associated with a ring
of emission at a radius of 60 to 95 AU.

Here we present the first high spatial resolution 24, 70 and 160 \um
images of Vega obtained with the {\it Spitzer Space Telescope}
\citep{werner04}. The sensitivity and resolution of {\it Spitzer}
allow us to detect fainter and more extended emission in the Vega
debris disk system than was known previously. Details about
observations, data reduction, and stellar photosphere subtraction are
presented in \S \ref{vega_obs}. The resolved images are discussed in
\S \ref{vega_morphologies}. The true nature of the disk can be
revealed by studying the radial profiles of the disk, which are
presented in \S \ref{vega_radialprofiles}, with interpretations of
disk temperature structures and masses in \S
\ref{vega_temp_structure}. The origin of the debris we detected with
{\it Spitzer} is discussed in \S \ref{vega_debris_origin}, followed by
our conclusions in \S \ref{vega_conclusion}.

\section{Observations, Data Reduction, and Photospheric Subtractions} 
\label{vega_obs}

Vega was imaged on 11 April 2004 with the Multiband Imaging Photometer
for {\it Spitzer} (MIPS; \citealt{rieke04a}). Multiple dithered 3 sec
exposures were obtained using the 24, 70 (both coarse and fine scales), and
160 \um channels. Details about the observations are listed in Table
\ref{obslog} along with the combined sensitivity at each band. 

The data were processed using the MIPS instrument team Data Analysis
Tool (DAT; \citealt{gordon04a,gordon04b}) for basic reduction (dark
subtraction, flat fielding/illumination correction). In addition, the
24 \um images were processed to remove a vertical ``jail-bar'' pattern
caused by the bright saturated central star. Due to the known
transient behaviors associated with the 70 \um detectors, a detector
dependent structure in the coarse scale mode data was removed by
subtracting column averages from each exposure with the source region
masked. Similar artifacts were taken out in the 70 \um fine mode data
by subtracting the off-source chopped background observations. The
processed exposures were then mosaicked together.

\begin{figure}
\figurenum{1}
\label{vega_radprof_a24} 
\plotone{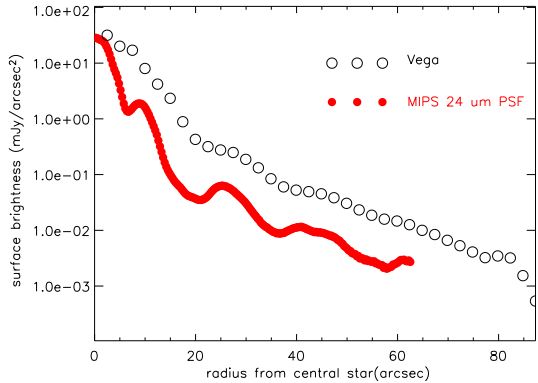}
\caption{Radial profile of Vega (photosphere $+$ disk) at 24 \um after
  background subtraction. 
For comparison, the radial profile of an observed point source is 
scaled to Vega's photosphere flux and shown as filled circles. 
The observed radial profile lies consistently above the point source
profile. 
The contrast between the bright and dark Airy rings in the PSF
signatures is not as prominent as the one in the point source,
suggesting the existence of a tenuous disk.}
\end{figure}

The Vega disk at 24 \um is tenuous compared to the very bright
photosphere (7.4 Jy). The existence of a very low surface brightness
disk is best illustrated by examining the radial profile (Figure
\ref{vega_radprof_a24}). Relative to the observed Point Spread
Function (PSF) scaled to Vega's photospheric level, the observed
radial profile lies consistently above the scaled point source
profile. The contrast between the bright and dark Airy rings in the
image of Vega is much lower, implying the brightness of the dark Airy
rings has been filled in by a low surface brightness, extended
component.

Vega's photosphere contributes a significant fraction of the flux near
the center of our images at both 24 and 70 \mm, while the emission
from Vega's debris disk dominates most of the flux at 160 \mm.  To
study the emission from Vega's debris disk the photospheric
contributions have to be removed in all three channels. To remove
Vega's photosphere, we subtracted reference observed PSFs, scaled to
Vega's photospheric flux, from our images. Changing the PSF scaling
factor by more than 10\% resulted in artifacts in the subtracted
images at the position of the core of the PSF and/or at the positions
of the Airy minima and maxima. We conclude that our photospheric
removal is accurate to 10\% within any given resolution element in the
resulting photosphere-subtracted images.

\begin{figure}
\figurenum{2}
\label{vegadisk_a24} 
\plotone{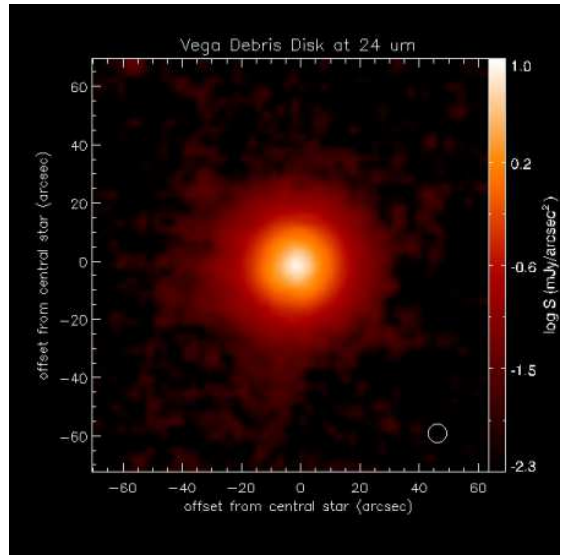}
\caption{ Vega disk at 24 \um displayed with logarithmic scaling. North is
  up and east is toward the left. Due to the saturation of the central
  star at 24 \mm, all the negative values after PSF subtraction have
  been excluded in the final mosaic, resulting in a smooth image at
  the core region (r $\leq$ 6\arcsec). The coverage (effective
exposure time) near the center is $\sim$half of that outside the
saturation region. The instrument beam size (FWHM) of 6\arcsec~at 24 \um
is shown as a white circle in the lower-right corner. }
\end{figure}

\begin{figure}
\figurenum{3a}
\label{vegadisk_a70a} 
\plotone{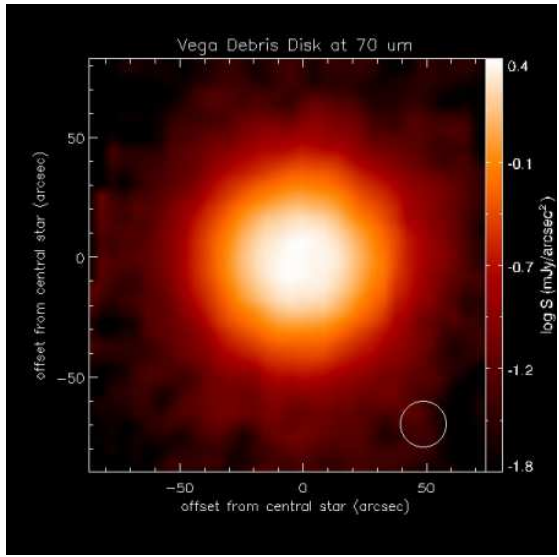}
\caption{Vega disk at 70 \um in coarse scale mode displayed with
  logarithmic scaling. North is up and east is toward the left. The Vega
photosphere has been subtracted off by scaling an observed PSF star.
The instrument beam size (FWHM) of 18\arcsec~at 70 \um
is shown as a white circle in the lower-right corner.}
\end{figure}

\begin{figure}
\figurenum{3b}
\label{vegadisk_a70b} 
\plotone{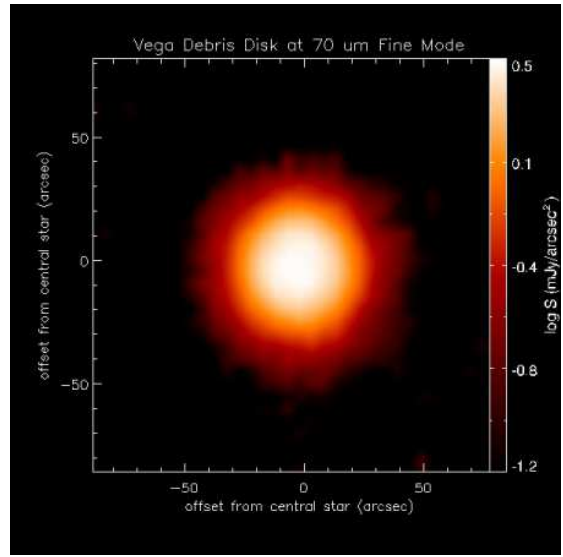}
\caption{Same as Figure \ref{vegadisk_a70a}, but in fine scale mode.}
\end{figure}

At 24 \um we processed the data for a PSF reference star (HD
217382) in exactly the same way as we did for Vega. Because the central
pixel in the Vega images was saturated, registering the reference PSF to
the Vega image was difficult. We took advantage of image latents present
in both data sets to do the registration and photosphere subtraction on
a frame-by-frame basis. We then mosaicked the resulting images to produce
the final 24 \um image of Vega's disk, shown in Figure \ref{vegadisk_a24}.
The mosaic does not show any saturation artifact at the center because
of the sub-pixel dithers used in the observation: a pixel that was
saturated in one image is partially overlapped by unsaturated neighboring
pixels in an image from a different dither position. There were enough
dither positions to provide partial coverage even at Vega's position
in the final mosaic. Because of this in-filling of the saturated area,
the effective exposure time in the central 6\arcsec~is about half
of that elsewhere.  Photospheric removal at 70 \um was more straight
forward. We registered the scaled reference PSF mosaic (HD 48915) to
the Vega mosaic by centroiding, and subtracted. The results are shown
in Figure \ref{vegadisk_a70a} (coarse scale) and \ref{vegadisk_a70b}
(fine scale).

\begin{figure}
\figurenum{4}
\label{vegadisk_a160} 
\plotone{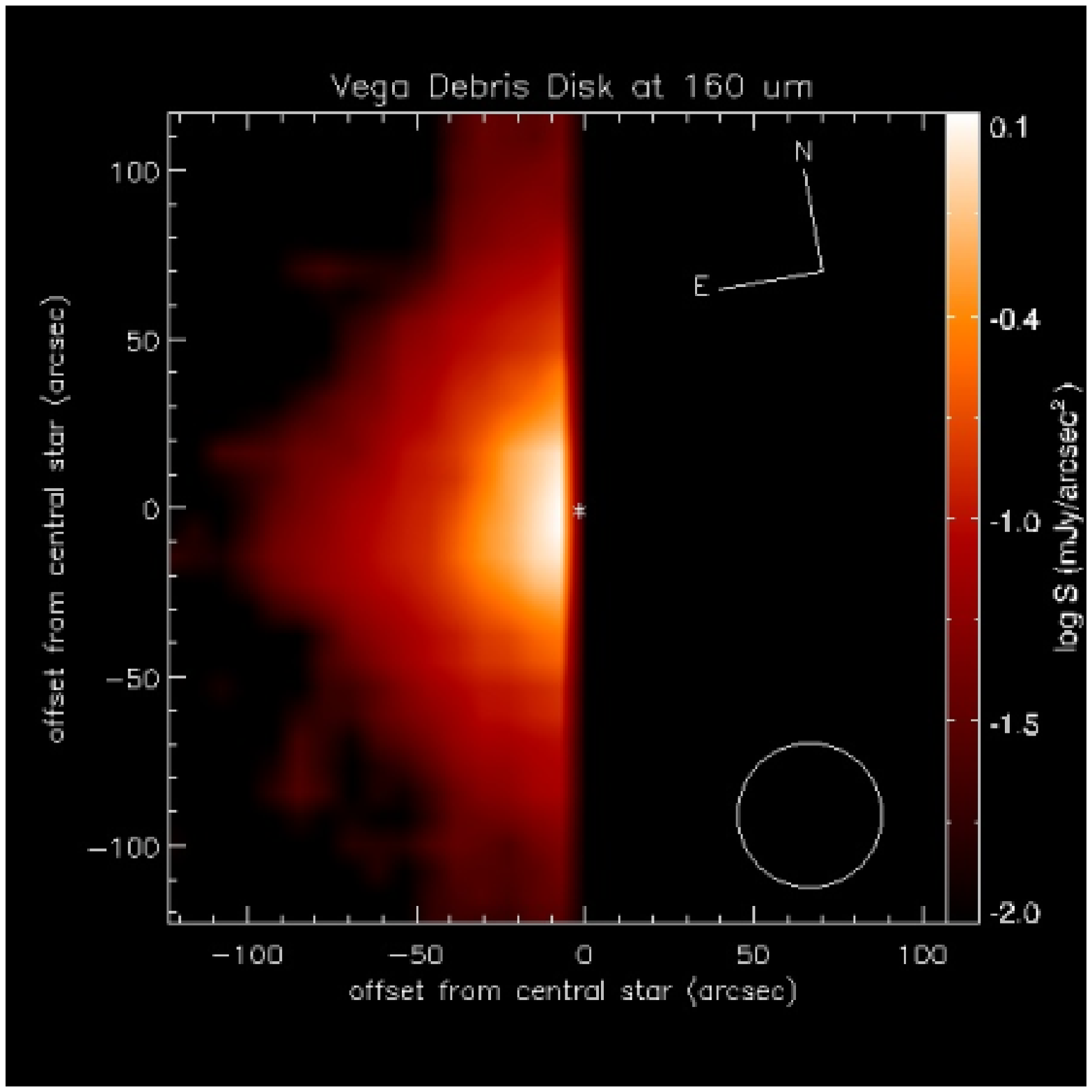}
\caption{Vega disk (half) at 160 \um displayed with logarithmic scaling. 
  The right side of the disk is contaminated by the
  spectral leak, and therefore not shown in the image. The white asterisk
at the center marks the stellar position. The instrument beam size (FWHM) of 40\arcsec~at 160 \um
is shown as a white circle in the lower-right corner.}
\end{figure}

The MIPS 160 \um array suffers from a spectral leak caused by an
internal reflection in the optical train allowing leakage from very
blue and bright Rayleigh-Jeans sources to contaminate the signals at
160 \mm. However,
the spectral leak image is offset to one side of the true 160 \um image, and
the brightness of the leak is proportional to the photospheric flux. 
Comparison of 160 \um images of stellar (blue) sources with images
of asteroids shows that the near-infrared leak contributes very little to the
160 \um images on the opposite side of the source location. The
predicted flux for Vega's photosphere is 162 mJy, which is much
fainter than the expected disk brightness at 160 \mm. 
We took advantage of this situation by using only the half of the 
160 \um image where the leak contribution is negligible.
We also subtracted a scaled (red) 160 \um reference PSF (asteroid
Harmonia) from the Vega mosaic, 
using the pointing information (accurate to $<$ 1\arcsec) to register. The result
is shown in Figure \ref{vegadisk_a160}.  

\section{Disk Morphologies at 24, 70 and 160 \mm}
\label{vega_morphologies}

We define our observed sensitivity based on the 1-$\sigma$ background
noise per pixel using blank sky area in the image. The 1-$\sigma$
background noise in the PSF-subtracted image is 11 ${\rm
\mu}$Jy/arcsec$^2$ at 24 \mm. The disk at 24 \um is symmetric and
centered at the star position; no obvious asymmetry is seen in the
image. At the 1-$\sigma$ level, the 24 \um disk extends to
$\sim$43\arcsec~(330 AU) in radius.  The total flux density (within
the 1-$\sigma$ contour) is $\sim$ 1.5 Jy $\pm 10\%$\footnote{10\%
including errors in absolute flux calibration, and in color correction
(less than 5\% for a blackbody temperature of 95 K).}.

This flux density value is in agreement with the {\it IRAS} 25 \um
measurement. The quoted {\it IRAS} 25 \um flux density for Vega is $\sim$10.5 Jy
(combining {\it IRAS} PSC and FSC). Based on Kurucz models, Vega's
photosphere is $\sim$6.63 Jy at 25 \mm. The relation between the {\it IRAS}
quoted flux and actual flux is $F_{actual} = F_{quoted} / K $, where
$K$ is the color correction factor. The color correction factor is
1.41 for T = 9750 K blackbody (Vega photosphere) and 0.83 for T = 95 K
(debris disk as determined by previous studies), suggesting that 
the actual flux from the disk is $\sim$ 1.4$\pm 10\%$ Jy at 25 \mm.

Vega's disk looks very similar in both 70 \um images. The fine scale
data were taken in 12 dithered positions with sub-pixel offsets to
provide a good spatial sampling for the disk. There is no obvious
clumpy structure in the fine scale image. Similar to the disk image at
24 \mm, the disk at 70 \um appears symmetric and smooth. At a
1-$\sigma$ detection limit, the outermost boundary of the disk at 70
\um (coarse scale mode) is $\sim$70\arcsec~(543 AU) in radius. The
total flux (within the 1-$\sigma$ contour) is $\sim$7 Jy ($\pm 20\%$).

Assuming, based on the 24 and 70 \um morphology, that the disk is also
azimuthally symmetric at 160 \mm, we are able to determine that the
outermost extent of the disk at 160 \um is $\sim$105\arcsec~(815 AU)
in radius (1-$\sigma$). Assuming the bad side of the disk image
(affected by the leak) has similar flux as the good side, the total
flux from the disk at 160 \um (within the 1-$\sigma$ contour) is
$\sim$4 Jy $\pm 20\%$.

Within the resolution errors, the disk image is circular at all wavelengths,
suggesting a face-on disk. This
behavior is consistent with Vega being a pole-on star
\citep{gulliver94}. Based on previous studies ({\it IRAS, ISO} and
SCUBA), the spectral energy distribution (SED) of the disk can be
characterized as a blackbody of T = 95 K with a 
$\lambda^{-1}$ emissivity. The fluxes measured by MIPS match the
previously observed SED within 15\%. 

A striking immediate result from the MIPS images is the physical
size of the Vega disk. Submillimeter and millimeter observations show a low surface
brightness ring with radius $\sim$11\arcsec, much 
smaller than the emission seen in the MIPS bands. In
contrast, MIPS images of the Fomalhaut debris disk \citep{stapelfeldt04} indicate that
the submillimeter and far-infrared morphologies agree well. 
A second surprise is the presence of material warm enough to be
detected in the MIPS bands at a large distance from the star.
\citet{artymowicz97} estimated the size of the disk if the grains are
blown out by radiation pressure, and suggested that the
disk could be up to a few thousand AU, consistent with our images.

\section{The Surface Brightness Radial Distribution} 
\label{vega_radialprofiles} 

\subsection{At 24 \um} 

\begin{figure}
\figurenum{5}
\label{rad24} 
\plotone{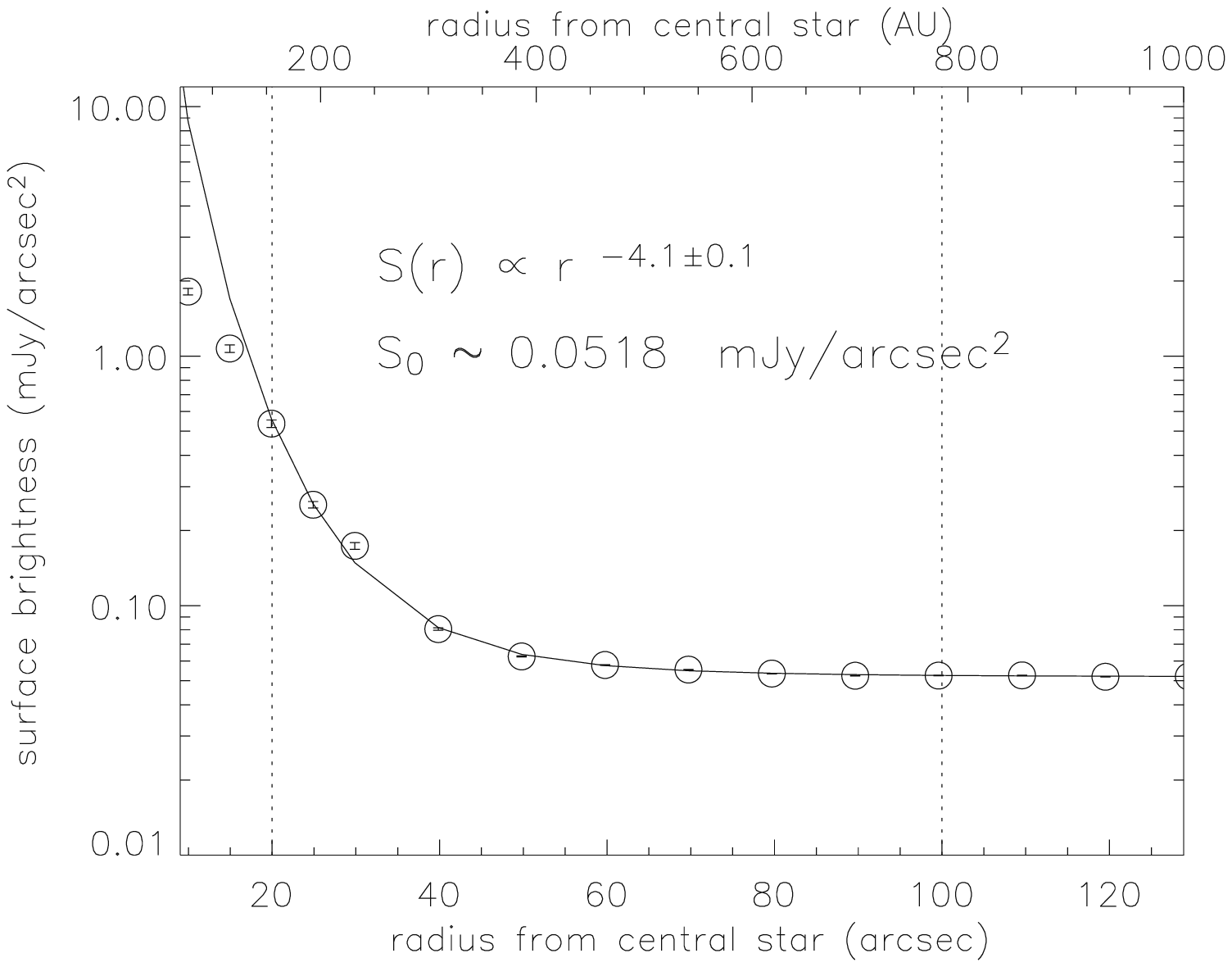}
\caption{Radial profile of the Vega disk at 24 \mm. A power law plus a
constant background are used to fit the data points between the two
dotted lines. The best-fit power law index is $-4.1\pm0.1$.}
\end{figure}

Because the disk is almost perfectly face-on, its extent can be best
shown in an azimuthally averaged radial intensity profile. The
advantage of using radial profiles is that the noise is reduced by
$\sqrt{N}$, where N is the total number of pixels used in averaging
the intensity at a given radius. The radial surface brightness profile
of the disk at 24 \um is shown in Figure \ref{rad24}. A bin size of
2\farcs5 (1 pixel) was used to calculate the average brightness at a
given radius {\it r} when $r \le$ 40\arcsec, but a bin size of
10\arcsec~(4 pixels) was used when $r >$ 40\arcsec~to increase the
signal-to-noise, up to 250\arcsec~($\sim$ 4\arcmin) from the star.  A
power law plus a constant background ($S(r) = S_o^{24} + r^{-\alpha}$)
was used to fit the data points between 22\arcsec~and 150\arcsec~to
determine the true sky background ($S_o^{24}$) and the flux radial
dependence.  A power index of $-$4.1$\pm$0.1 is indicated, based on a
least-squares fit.

\begin{figure}
\figurenum{6}
\label{rp24} 
\plotone{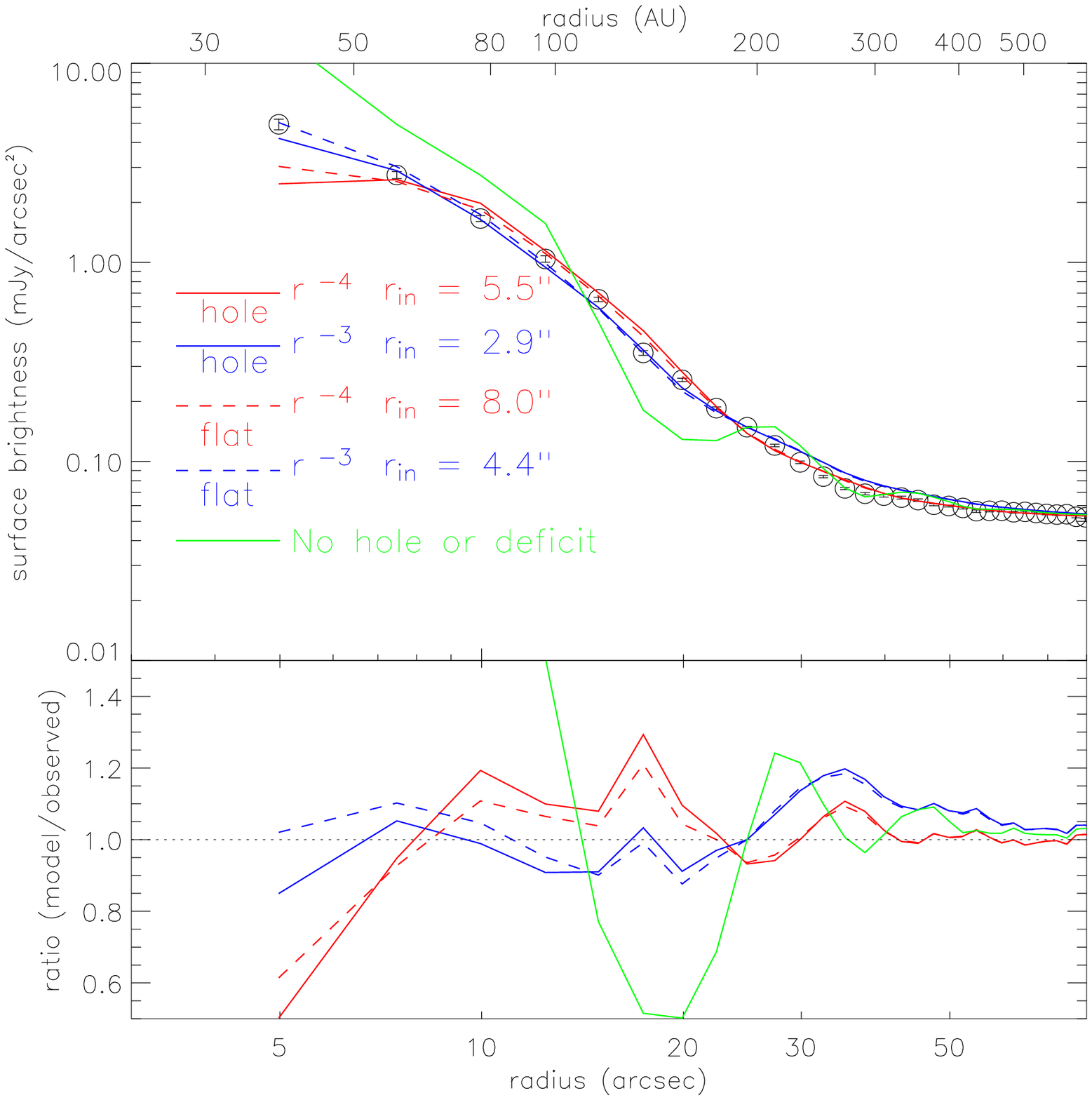}
\caption{Radial profile of the Vega disk at 24 \um (open circles) 
  compared with modeled surface brightness distributions 
  after convolution with the instrumental beam. Note that the
  background value determined in Figure \ref{rad24} has been added
  back in the fits. The model fit with a steep power law (-4 or -3)
  and no inner boundary of the disk is shown as a green solid line, 
  which resembles a PSF. The other models
  are: Model A -- an empty hole with $r^{-4}$ as a solid red line; Model
  B -- an empty hole with $r^{-3}$ as a solid blue line; Model A' -- a
  flat distribution with $r^{-4}$ as a dashed red line; Model B' -- a
  flat distribution with $r^{-3}$ as a dashed blue line. 
  The lower panel of the plot shows the ratios between the modeled 
  and the observed radial profiles.}
\end{figure}

The dashed lines on Figure \ref{rad24} mark the data range for the
power law fitting. The power index of $-$4.1 fits fairly well for the
disk surface brightness when $r >$ 200 AU, but deviates significantly
when $r <$ 150 AU. To understand the inner structure of the disk, we
first compute a disk model with a $r^{-\alpha}$ surface brightness
distribution and then convolve the model disk with the 24 \um
PSF. With a rapid power law fall-off ($\alpha =$4), the radial profile
of the convolved model disk (shown as a green solid line in Figure
\ref{rp24}) is similar to the point source PSF, which was not seen in
the observed data. This comparison suggests the surface brightness
distribution must flatten or go to zero (a real physical empty hole)
in the inner part of the disk.

We first investigated the possibility of a real empty hole in the
surface brightness distribution. By adjusting power-law indices and
inner hole sizes, we are able to reproduce the observed radial profile
using two different models. Model A, which is $r^{-4}+S_o^{24}$ with
an inner disk hole ($r_{in}$) at 5.5\arcsec~with an error of $\pm$7\%
at 90\% $\chi^2$ confidence level (shown as a red solid line in Figure
\ref{rp24}), fits very well when $r \ge$ 200 AU as expected.  Model B,
which is $r^{-3}+S_o^{24}$ with $r_{in}$ = 2.9\arcsec~with an error of
$\pm$17\% (shown as a blue solid line in Figure \ref{rp24}), fits very
well when $r <$ 200 AU, but has slightly higher surface brightness at
$r >$ 200 AU compared to the $r^{-4}$ power law. The lower panel of
Figure \ref{rp24} shows the ratio between the model and observed
radial profiles.

We have shown that a real empty hole in the surface brightness profile
can fit the data well. Next we investigated the possibility of a flat
surface brightness distribution in the inner part of the
disk. Assuming the surface brightness is constant, $S(r_{in}$), when
$r \le r_{in}$, rather than zero in the empty hole case, we found two
models that can fit the observed profile. Model A', which is
$r^{-4}+S_o^{24}$ with an $r_{in}$ = 8.0\arcsec~with an error of
$\pm$5\% (shown as a red dashed line in Figure \ref{rp24}), fits very
well when $r \ge$ 200 AU.  Model B', which is $r^{-3}+S_o^{24}$ with
$r_{in}$= 4.4\arcsec~with an error of $\pm$14\% (shown as a blue
dashed line in Figure \ref{rp24}), fits very well when $r <$ 200 AU,
but has slightly higher surface brightness at $r >$ 200 AU compared to
the $r^{-4}$ power law. The deficit in the flat surface brightness
distribution is very large ($\ge$99\%) when compared to the
extrapolation of the steep power law to the disk center. The
difference between an empty hole and a flat distribution in terms of
surface brightness is insignificant. For $r >$ 200 AU, Model A and A'
provide a good fit to the data whereas Model B and B' match the data
well for $r <$ 200 AU. The most important conclusion from these
fittings is that the disk surface brightness distribution follows
simple power-law dependences, implying that the disk density structure
is simple and smooth.

\subsection{At 70 \um} 

\begin{figure}
\figurenum{7}
\label{rad70}
\plotone{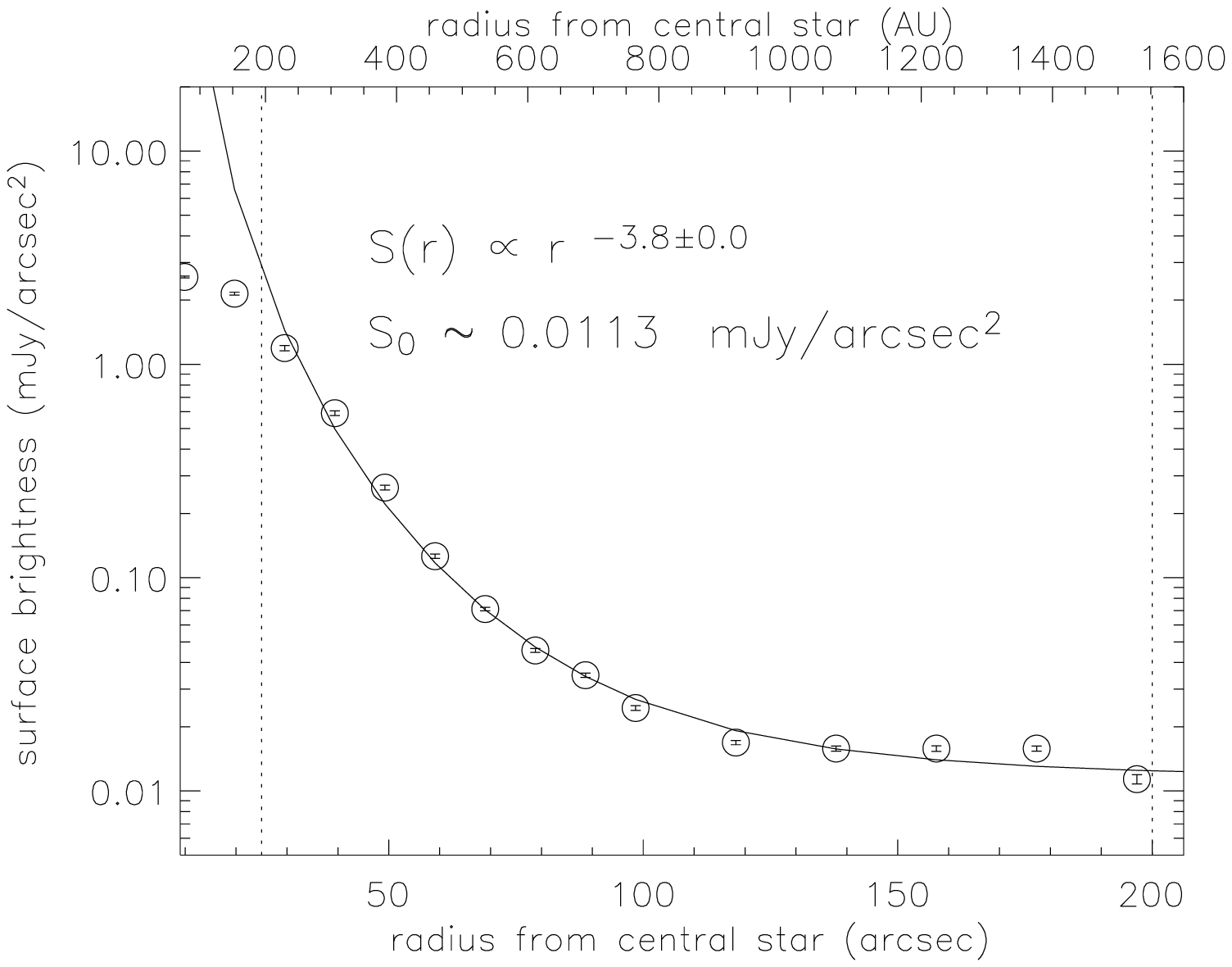}
\caption{Radial profile of the Vega disk at 70 \mm. A power law plus a
constant background are used to fit the data points between the two
dotted lines. The distribution is found to be consistent with 
a $r^{-3.8}$ power law.  }
\end{figure}

\begin{figure}
\figurenum{8}
\label{rp70}
\plotone{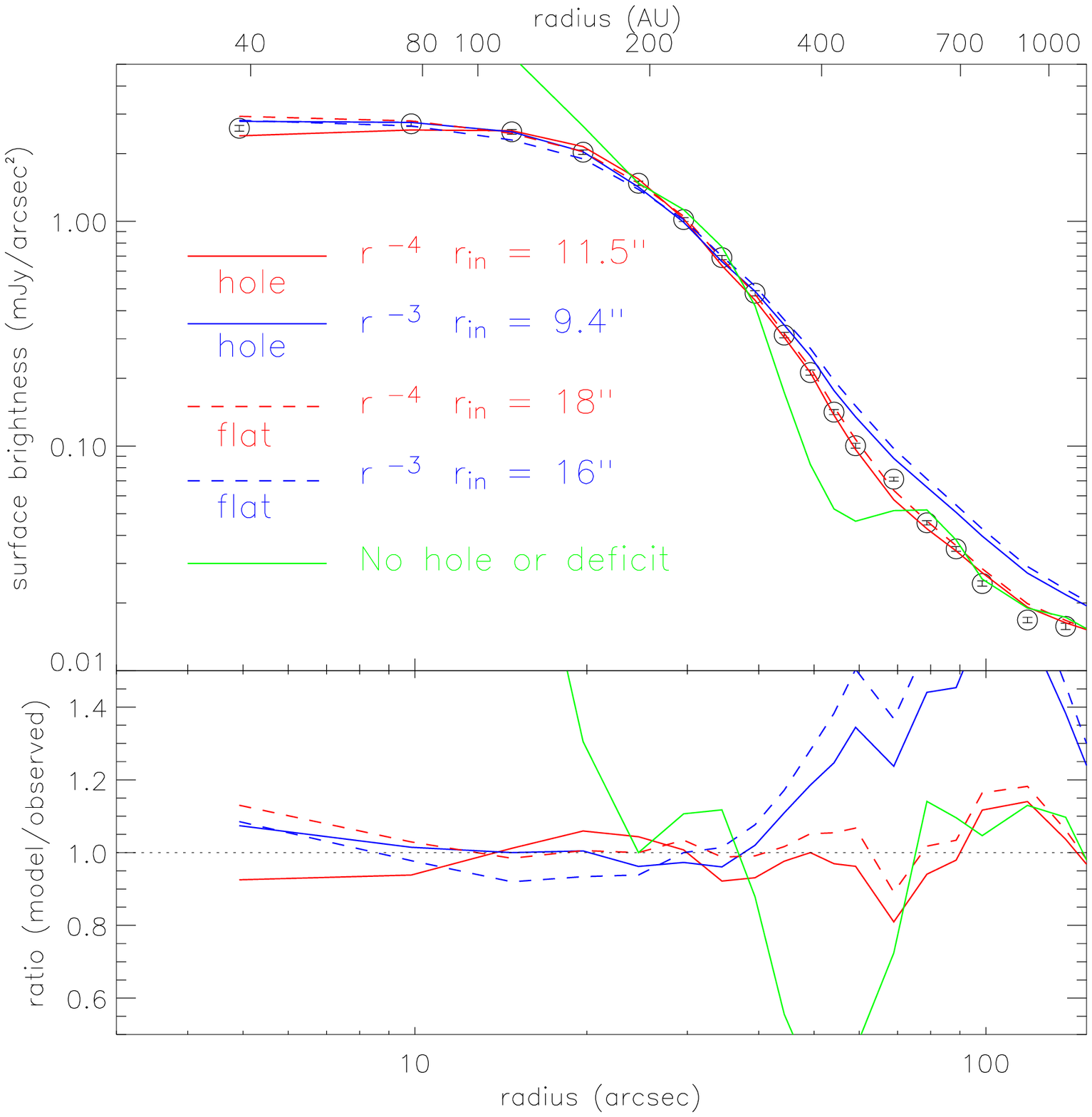}
\caption{Radial profile of the Vega disk at 70 \um compared with model
surface brightness distributions after convolution with the
  instrumental beam. Similar to Figure \ref{rp24}, the background
  value has been added back in the model fits. 
  Symbols and color scheme are the same as Figure \ref{rp24}. 
  The lower panel of the plot shows the ratios between the modeled 
  and the observed radial profiles. }
\end{figure}

The 70 \um radial profile of the Vega disk was computed using an
average of the data from both observing modes for $r \le$40\arcsec,
but only the coarse scale mode data when $r >$ 40\arcsec. The combined
radial profile is shown in Figure \ref{rad70}. Fitting with a power
law plus a constant background function, $S(r) = S_o^{70} +
r^{-\alpha}$, the surface brightness of the disk shows a $r^{-3.8}$
power law for $r >$ 200 AU with a very small error in the
index. Similar approaches to those pursued at 24 \um (hole or flat
distribution) were used to fit the inner part of the radial profile at
70 \mm. The results are shown in Figure \ref{rp70}. For an empty hole
in the surface brightness distribution, Model A gives a $r_{in}$ =
11.5\arcsec~with an error of 5\% and Model B give a $r_{in}$ =
9.4\arcsec~with an error of 7\%. For a flat distribution, Model A'
gives a $r_{in}$ = 18\arcsec~with an error of 6\% and Model B' gives a
$r_{in}$ = 16\arcsec~with an error of 6\%, and the deficit is larger
than 99\% when compared to the extrapolation of the power law to the
disk center. For the inner part of the disk ($r \le$ 350 AU), Model
A/A' and Model B/B' give similar results, but Model A and A' give a
much better fit for $r > $350 AU.

\subsection{{\it [24]$-$[70]} color distribution}

\begin{figure}
\figurenum{9}
\label{cc24_70}
\plotone{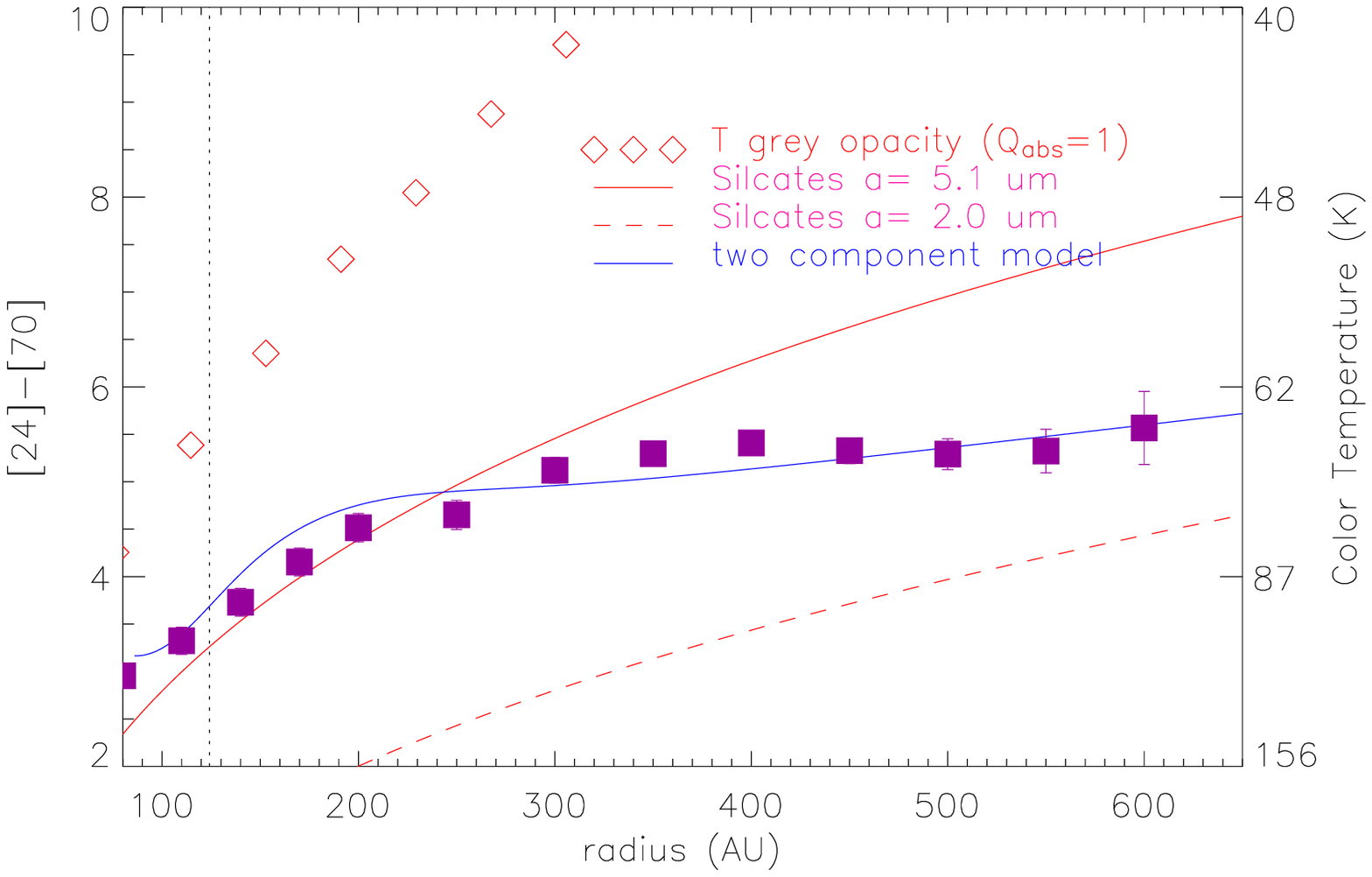}
\caption{Radial dependent {\it [24]$-$[70]} distribution for the Vega
disk. The observed color is plotted in filled squares. The color based
on the canonical dust temperature is plotted as open diamonds,
compared to the ones derived from single sized silicates ($a=$ 5.1 \um
as the solid red line and $a=$ 2 \um as the dashed red line). The
{\it [24]$-$[70]} color derived from our two-component model (see \S
\ref{preferred}) is also plotted as the blue dashed line.}
\end{figure}

To compare the disk structure between 24 and 70 \mm, we first
convolved the 24 \um disk image with a Gaussion kernel to match the 70
\um resolution. The radially dependent {\it [24]$-$[70]} color is
estimated at a given radius from the star by computing the flux ratio
between 70 and 24 \mm. The resultant radial color distribution is
shown in Figure \ref{cc24_70}. The {\it [24]$-$[70]} color initially
increases (color temperature decreases) with distance from the star,
but levels off between 300 and 600 AU (with a color temperature of
$\sim$67 K).  For comparison, we also overplot the expected color from
the canonical temperature distribution, $T_g = T_{\ast} (R_{\ast}/2
r)^{0.5} $ with $T_{\ast}=9750 K$ and $R_{\ast}=2.5 R_{\sun}$ for
Vega, assuming spherical blackbody radiators (i.e., absorption
efficiency $Q_{abs}=$ 1).

We compute the radial color values for single-size silicates assuming
the grains are heated only by stellar radiation and emit in an
optically thin case (for details, see section
\ref{vega_temp_structure}). As shown in Figure \ref{cc24_70}, the {\it
[24]$-$[70]} color in the inner part ($r <$ 200 AU) of the disk is
consistent with the emission from single-size silicates with radius
$a=$ 5.1 \mm. Furthermore, a disk with only 2 \um grains does not
reproduce the observed color distribution at all. The temperature
structure of the disk cannot be easily explained by thermal emission
from a disk composed of single-size grains. To produce the observed
{\it [24]$-$[70]} color, additional components must be invoked.

\subsection{At 160 \um} 

Since the disk is bright and extended at 160 \mm, we were able to
extract a disk radial profile based on the good half of the disk
image. To verify our 160 \um radial profile is legitimate we also
extracted radial profiles of a true (no leak) 160 \um source (asteroid
Harmonia) and a calibration star (HD 3712, K0IIIa) only using the good
half of the image. We found that the radial profiles extracted only
from the good half the images look similar, suggesting the good half
of the image is not affected by the leak.

\begin{figure}
\figurenum{10}
\label{rp160}
\plotone{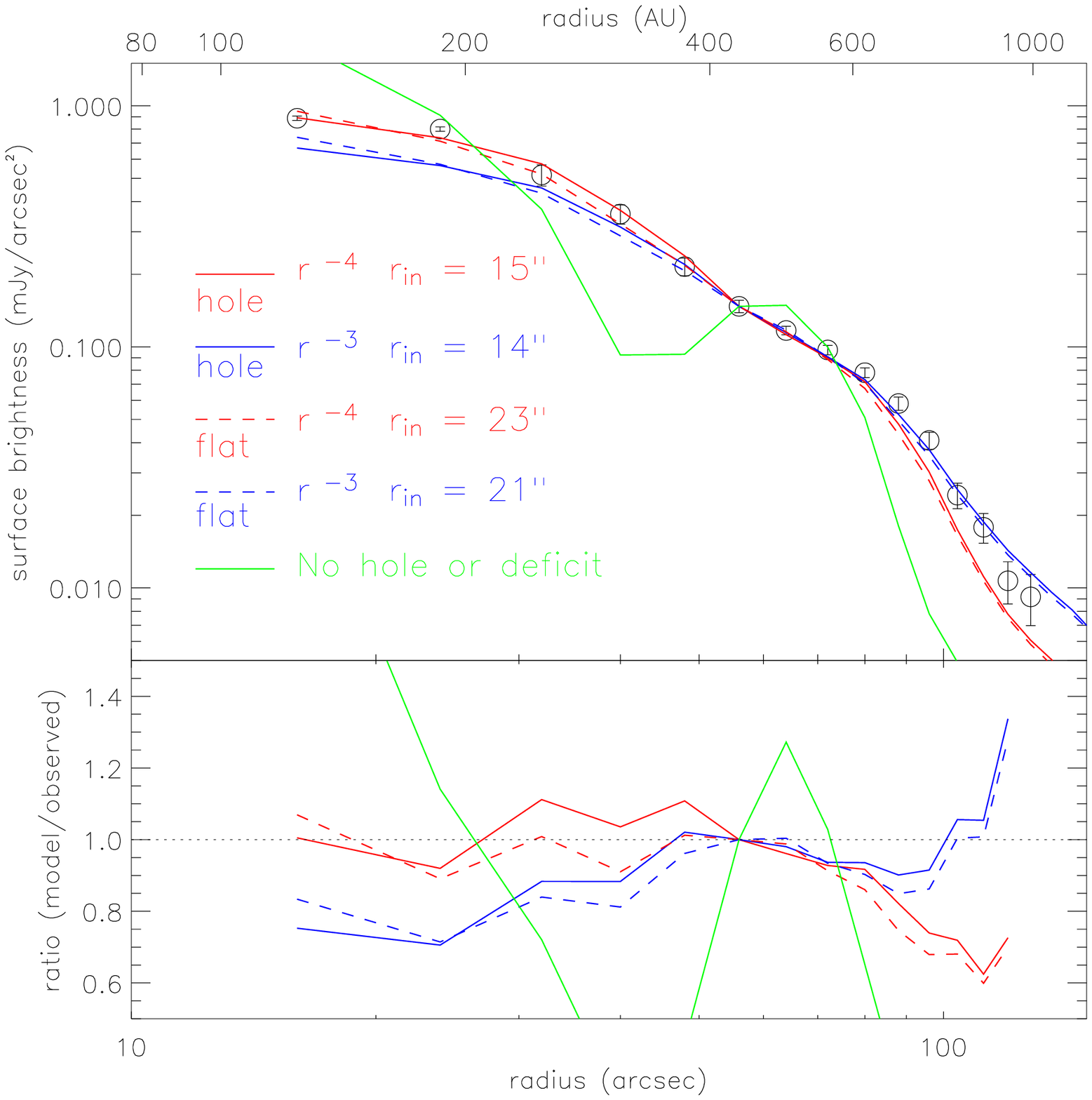}
\caption{Radial profile of the Vega disk at 160 \um compared with model
surface brightness distributions after convolution with the
  instrumental beam. The background value is not included in the
  fits. Symbols and color scheme are the same as Figure \ref{rp24} and
  \ref{rp70}.  The lower panel of the plot shows the ratios between
  the modeled and the observed radial profiles.}
\end{figure}

The Vega disk radial profile at 160 \um is shown in Figure \ref{rp160}
after background subtraction. Similar approaches (hole or flat
distribution) were used to fit the radial profile. The results are:
Model A (hole, r$^{-4}$) gives a $r_{in}$ = 15\arcsec; Model A' (flat,
r$^{-4}$) gives a $r_{in}$ = 23\arcsec; Model B (hole, r$^{-3}$) gives
a $r_{in}$ = 14\arcsec; Model B' (flat, $r^{-3}$) give a $r_{in}$ =
21\arcsec. The errors in these model fits are on the order of
30\%. Again, the deficit in the flat distribution is greater than 99\%
when compared to the extrapolation of the steep power law to the disk
center.

\subsection{Summary}

The surface brightness distribution at the 24 \um band agrees with a
radial-dependent power law with index of $-$3 (inner part) or $-$4
(outer part). At 70 \mm, the $r^{-4}$ power law fits better than the
$r^{-3}$ in the outer part of the disk. In addition, this
radial-dependent power law cannot apply all the way to the star; a
hole or a flat distribution in the disk is required. The inner edge of
the power-law surface brightness distribution depends on whether the
central region contains no flux or a constant non-zero flux. In the
latter case, the inner disk radius is $\sim$50\% larger. However, the
current data are not adequate to distinguish between a empty hole and
a flat non-zero central flux deficit. Therefore, the inner boundary of
the disk is determined as a weighted average of these radii obtained
from different fits in all three bands. We conclude that the inner
boundary of the Vega debris disk has a radius of
11\arcsec$\pm$2\arcsec~(or 86$\pm$16 AU). The ``hole'' is not required
to be physically empty; rather it is a region lacking material that
emits strongly at 24, 70, and 160 \mm. Interestingly, the radius of
the 11\arcsec$\pm$2\arcsec~hole is about the same size as the 850 \um
(and mm-wave) ring. This implies that the material we detect in the
MIPS bands might originate from the submillimeter ring (large parent
bodies) where collisional grinding generates small debris that is
blown out by radiation pressure to the distances where we detect their
emission.

\section{Disk Structure and Mass} 
\label{vega_temp_structure} 

For particles moving in a steady-state flow, the mass gain/loss rate
$\frac{\vartriangle M}{v \vartriangle t}$ through any annulus is
expected to be constant. If the particles are moving at their terminal
velocities ($v \sim$ const.), the surface number density ($\Sigma$) is
expected to be proportional to $r^{-1}$ due to mass conservation
($\vartriangle$$M = 2 \pi r v$$\vartriangle$$r \Sigma \sim$
const.). Therefore, for a radiation pressure driven outflow, the
surface particle number density should follow an inverse radial power
law if the particles have reached their terminal velocities. In \S
\ref{preferred}, we will show the Vega debris disk is well modeled
with an $r^{-1}$ surface number density power law.

\subsection{Preferred Model}
\label{preferred}

We assume that the disk is axially symmetric and geometrically
thin, viewed face-on, 
and that the surface number density is governed by a simple radial
power law; i.e., $\Sigma(r) = \Sigma_0 (r/r_0)^{-p} $ where $r_0$ is
the inner radius of the disk. Based on results in \S
\ref{vega_radialprofiles}, we set $r_0=$ 86 AU and simplify the disk
inside $r_0$ as a physically empty space. 
The emission from grains at a given
radius $r$ and wavelength $\lambda$ can be written as 
\begin{equation} 
 dF_{\lambda} = dn(r) \frac{\pi a^2}{d^2} Q(a,\lambda)
B_{\lambda}(T_r), 
\end{equation} 
where $d$ is the distance to Vega and $dn(r) = 2\pi\Sigma(r) r dr$ is
the total number of grains of a given radius between $r$ and $r+dr$. 

\begin{figure}
\figurenum{11}
\label{grains_temp}
\plotone{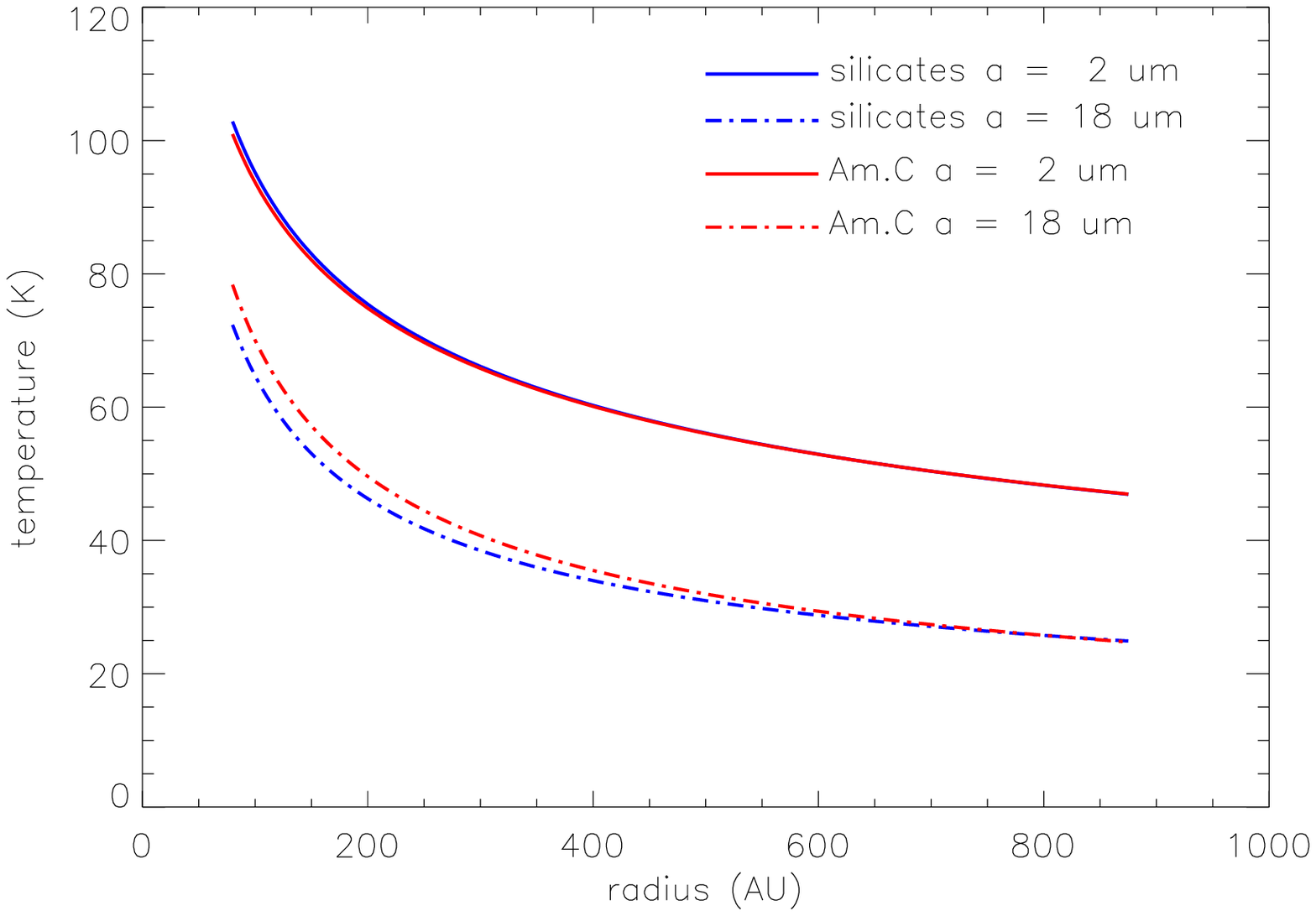}
\caption{Thermal equilibrium temperature distribution for different
  grains in the Vega environment. Astronomical silicates are plotted in blue,
  and amorphous carbon grains in red. Solid lines are for
  small grains ($a=$ 2 \mm) and dashed-dotted lines are for large grains
($a=$ 18 \mm).}
\end{figure}

To probe the nature of the Vega disk further, we have fitted it with
models including realistic grain optical properties. A dust grain
temperature for a given set of grain parameters (composition, grain
radius $a$, i.e. $Q_{abs}(a,\lambda)$) at a given wavelength
($\lambda$) and location from the star ($r$), can be computed assuming
the dust is only heated by the star. We use the photospheric flux from
the Kurucz model for Vega as the input radiation field. The absorption
coefficient is calculated with Mie theory, using the optical constants
for astronomical silicates ($\rho_g$= 3.5 g/cm$^3$,
\citealt{laor93}). The grain temperatures at different radii from the
star, $T(r)$, shown in Figure \ref{grains_temp}, are computed based 
on balancing the energy between the absorption and emission by the 
dust grains (scattering can be ignored for the wavelengths of interest).

The total flux from an annulus ($r'$ to $r''$) by integrating Equ.~(1)
is 
\begin{equation}
F_{\lambda}(r') = 2 \pi \Sigma_0 \frac{\pi a^2}{d^2} Q(a,\lambda) r_0^p
\int_{r'}^{r''} r^{-p+1} B_{\lambda}(T_{r}) dr ~.
\end{equation}
If the annulus is small enough (i.e., $r' \sim r''$ ), the flux can be
written as 
\begin{equation}
F_{\lambda}(r') = 2 \pi \Sigma_0 \frac{\pi a^2}{d^2} Q(a,\lambda) r_0^p
~(r')^{-p+1} B_{\lambda}(T_{r'}) (r'' - r') ~. 
\end{equation}
The surface brightness distribution, $S_{\lambda}(r')$, is the total
flux divided by the annulus area, i.e.,
\begin{equation}
S_{\lambda}(r') = 2 \Sigma_0 \frac{\pi a^2}{d^2} Q(a,\lambda) r_0^p
~(r')^{-p+1} B_{\lambda}(T_{r'}) \frac{r''-r'}{{r''}^2 - {r'}^2} ~.
\end{equation}

Equ.~(4) can be used to qualitatively evaluate the expected power
index in the surface number density.  For a given wavelength and set
of grain properties, the observed surface brightness, $S(r)$, is
proportional to $B_{\lambda}(T_{r}) \cdot r^{-p}$. In general, the
temperature, $T_{r}$, is proportional to $r^{-0.33}$ for small grains,
or $r^{-0.5}$ for large grains. Using the temperature distribution for
silicate grains with $a=$ 2 \mm, the Planck function at 24 \um
basically follows $r^{-2}$ in the inner part of the disk, but $r^{-3}$
in the outer part of the disk. In other words, if the observed 24 \um
flux is dominated by the emission from small grains, a $p=$ 1 power
index for the surface density distribution is preferred based on the
fits in \S \ref{vega_radialprofiles} ($r^{-3}$ for inner disk but
$r^{-4}$ for outer disk).

For proper comparison to the observed surface brightnesses (after
subtraction of the true sky background determined from the fits in \S
\ref{vega_radialprofiles}), the modeled surface brightness profile
using Equ.~(4) is then convolved with smoothed theoretical
PSFs\footnote{The observed PSFs are found to match well with smoothed
theoretical STinyTim PSFs. If the object is well resolved compared to
the instrument beam, convolving with a gaussian with a FWHM equal to
the instrumental resolution is sufficient. However, if the object is
not very well resolved, convolving with a gaussian will not match the
observed data since MIPS PSFs are not gaussian-like especially at 160
\mm.} to match the observed resolutions: 6\arcsec, 18\arcsec~and
40\arcsec~for 24, 70 and 160 \mm, respectively. All the fittings shown
below are restricted to the data points between 100 to 700 AU where
high signal-to-noise data are available. We first attempted to fit the
radial profiles using a single grain size, but could not obtain a
reasonable fit simultaneously for all three wavelengths. In fact, the
{\it [24]$-$[70]} color indicates that the disk is composed of a hot
component that emits efficiently at 24
\mm; while the comparable disk size at 70 and 160 \um indicates that the
disk is also composed of a cold component that emits efficiently at
longer wavelengths.

\begin{figure}
\figurenum{12}
\label{radprof_3bands}
\plotone{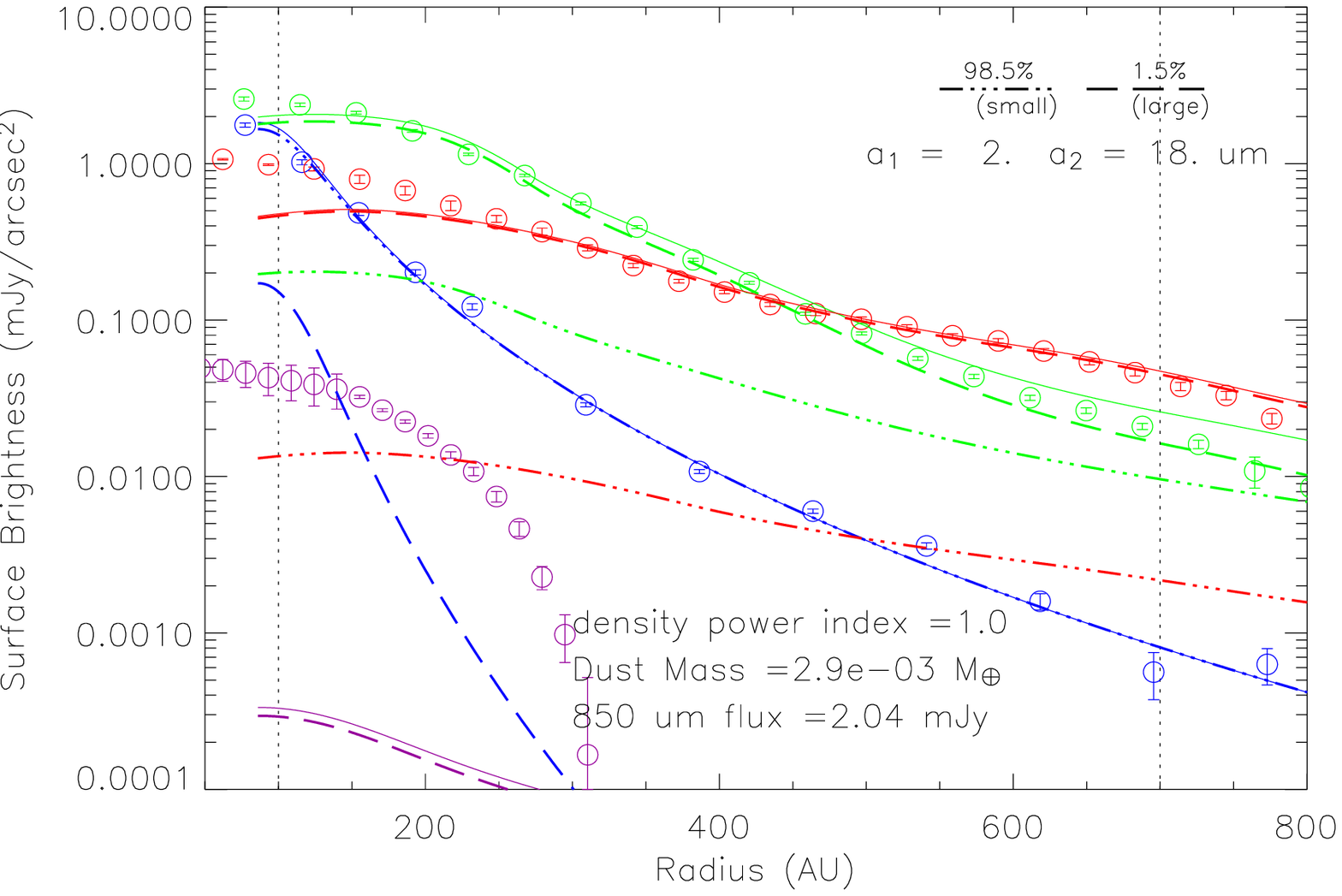}
\caption{The observed radial surface brightness profiles for Vega
  debris disk are plotted as open circles with colors representing
  different bands (24 \mm: blue, 70 \mm: green, and 160 \mm: red). 
  Model profiles from the large grain ($a=$ 18 \mm) component
  and the small grain ($a=$ 2 \mm) component are plotted as dashed lines and
  dashed-dotted lines, respectively, with colors indicating the different
  bands. The sum of the two components is plotted as solid lines
  with colors representing the different bands. 
  The 850 \um emission from the two-component 
  grain model is also plotted as a purple solid line for comparison
  to data from JCMT/SCUBA (purple open circles). 
}
\end{figure}

We are able to find a reasonable fit for all three radial profiles by
assuming the disk has an inverse radial surface density distribution
($p = 1$), and is composed of grains with sizes: $a_1$ = 2 \um and
$a_2=$ 18 \mm, with the former size accounting for 98.5\% of the disk
particles by number. The fits are shown in Figure
\ref{radprof_3bands}. Assuming the disk extends to 1000 AU, the total
mass of the material that emits at 24, 70 and 160
\um is 2.9$\times$10$^{-3}$ $M_{\earth}$, with 92\% of the mass from
the large grain component. However, the contributions of the small and
large components to the total grain surface area are comparable. 
Note that the poor quality of the fit at the central part of the 160
\um profile is not sensitive to variation of the model parameters. The
deficit at the central 160 \um model profile might suggest the
existence of a unresolved cold component close to the star. To further
constrain these parameters and estimate their uncertainties, we
searched large regions of parameter space, computing a $\chi^2$
statistic at each point. The search covered the following
four-dimensional hypervolume: for the surface density power index, $p$,
we searched from 0.5 to 3.0 with an interval of 0.1; for the number
fraction of the small grains ($f_1$), we searched from 10\% to 99.9\%
with logarithmic spacing; for the grain size in the component 1 (small
grains, $a_1$) and component 2 (large grains, $a_2$), we searched from
0.3 to 52 \um with logarithmic spacing. At 90\% confidence, we
conclude that $p = 1.0 \pm 0.2$, $f1 = 98^{+0.8}_{-3.2}$ \%, $a_1 =
2.0 \pm 0.7$ \mm, and $a_2 = 18^{+12}_{-6}$ \mm.  The total mass
estimated from the $\chi^2$ fitting is not well constrained
(2.9$^{+20}_{-2}\times$10$^{-3}~M_{\earth}$) since the
mass estimate is very sensitive to the specific size of the large
component.

\begin{figure}
\figurenum{13}
\label{radprof_powersize}
\plotone{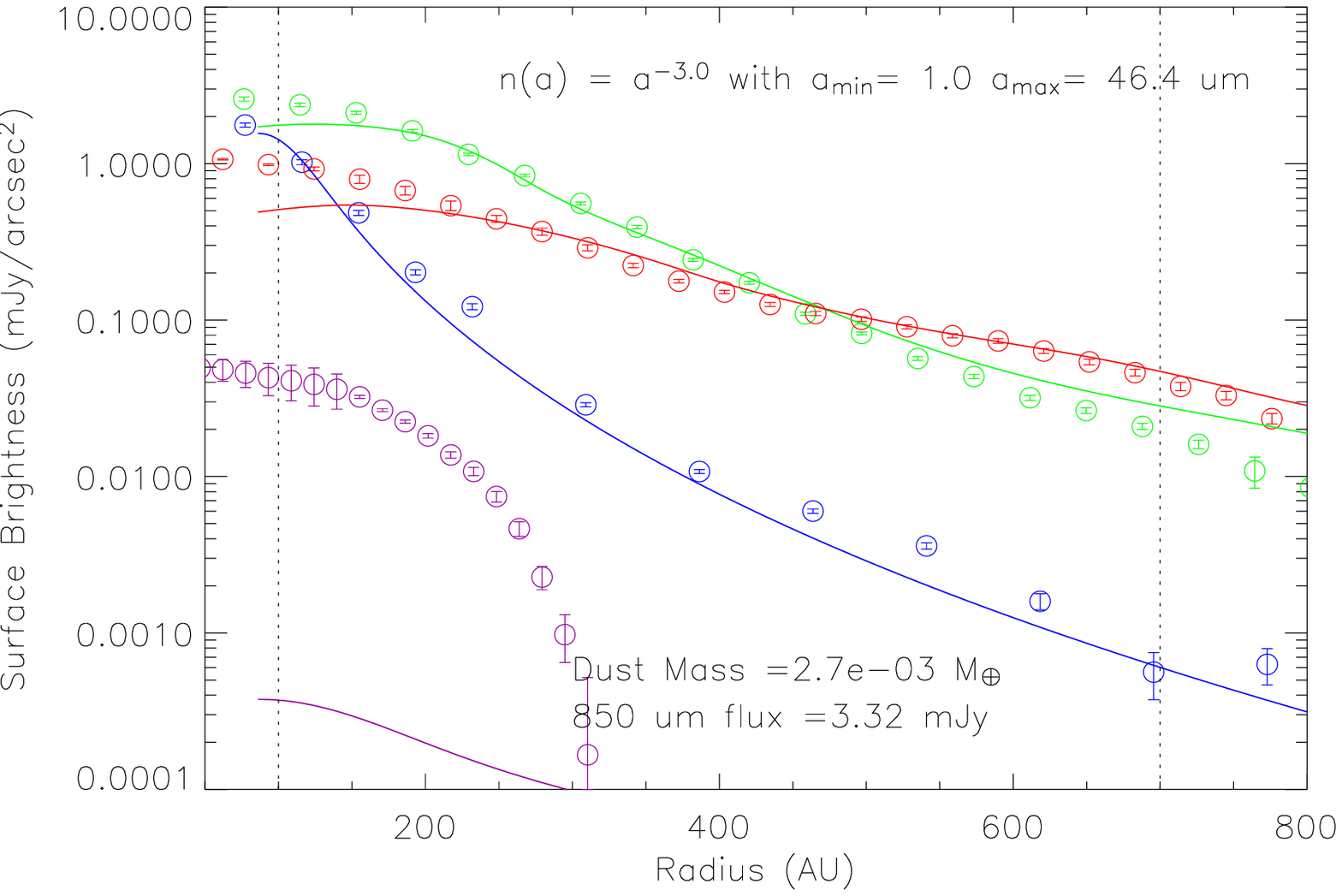}
\caption{Similar to Figure \ref{radprof_3bands}, but using the 
  model profiles from a grain size distribution: $n(a) \propto
  a^{-3}$ with a minimum size cut-off $a_{min}=$ 1.0 \um and maximum
  size cut-off  $a_{max}=$ 46 \mm. The symbols and lines are the same as in Figure
  \ref{radprof_3bands}.} 
\end{figure}

We also fit the radial profiles using a grain size distribution, $n(a)
\propto a^q$, with minimum and maximum size cut-offs, $a_{min}$ and
$a_{max}$. The parameter space searched for the best fit encompassed
values of $q$ ranging from -1.0 to -4.0, $a_{min}$ from 0.5 to 3.0
\mm, and $a_{max}$ from 10 to 60 \mm. We fixed the surface number
density index, $p$, at 1 because we found that the model profiles drop
too quickly for $p >$1 in the previous two-component $\chi^2$ fitting.
At the 90\% confidence level, we found a best fit with $q=-3.0\pm$0.5,
$a_{min} = 1.0^{+1.3}_{-0.3}$ \mm, and $a_{max} = 46 \pm 11$ \mm. The
best fit is shown in Figure \ref{radprof_powersize}.  With these
parameters, the total dust mass detected by MIPS is
2.8$\pm$1.1$\times$10$^{-3}$ $M_{\earth}$.

Our two-size and power-law distribution models both fit the surface
brightness profiles well at all three MIPS bands and can be used to
predict the surface brightness profiles at 850 \mm. The modeled 850
\um profile is shown as a purple line in Figures \ref{radprof_3bands}
and \ref{radprof_powersize}, after convolution with a Gaussian
beam-size of 16\arcsec. For comparison, the reprocessed 1998 SCUBA
archival data are also plotted as purple open circles after
subtracting a $\sim$5 mJy point source (representing the stellar
photosphere) and smoothing with an additional Gaussian beam-size of
7\arcsec. The total modeled flux emitted at 850 \um based on the two
models is less than 3.5 mJy, much lower than what was found in the
SCUBA data, 91$\pm$8 mJy \citep{holland05}.

\begin{figure}
\figurenum{14}
\label{radprof_powersize_amc}
\plotone{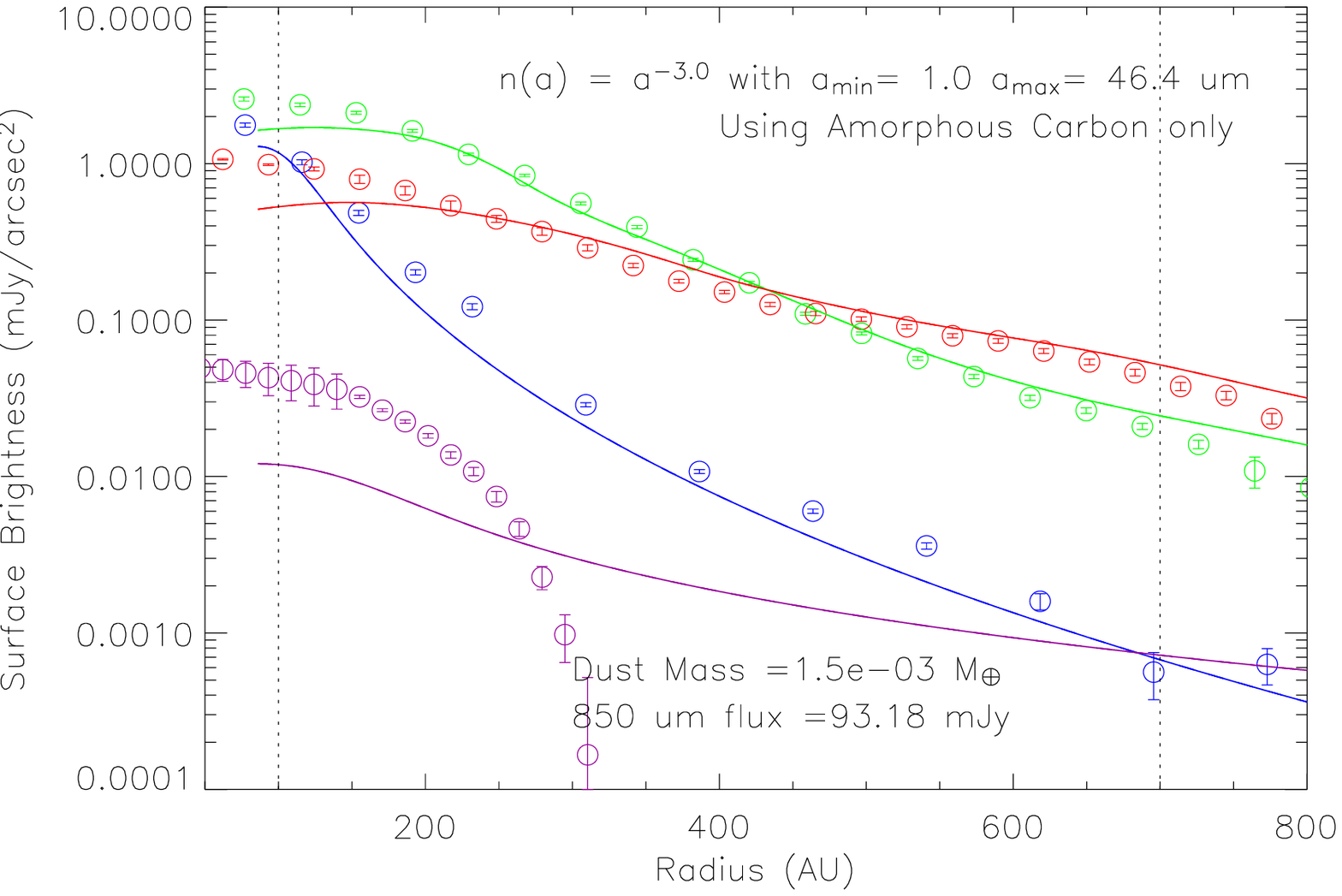}
\caption{Same as Figure \ref{radprof_powersize} but using amorphous
  carbon grains instead of silicates. At the MIPS wavelengths, the
  modeled profiles are very similar to the ones using silicate grains
  due to the similar temperature distributions and emission
  efficiency. However, for a particle with $a=$ 2 \mm, the emission 
  efficiency ($Q_{abs}$) at 850 \um for amorphous carbon grains is
  twice larger than the one for silicates. As a result, the modeled
  emission at 850 \um using amorphous grains is much brighter and more
  extended compared to the emission using silicates. 
} 
\end{figure}

To test the sensitivity of the model fits to the assumption of
astronomical silicates, we also fit the radial profiles using
amorphous carbon grains ($\rho=$ 1.85 g/cm$^3$,
\citealt{zubko96}). Similar to the silicate grains, a temperature
distribution is computed for each of the grain sizes.  Without
changing the grain size distribution, a satisfactory fit is
automatically obtained at all three MIPS wavelengths (Figure
\ref{radprof_powersize_amc}). This is because the resultant
temperature distributions in this environment are very similar between
silicates and amorphous carbons (Figure \ref{grains_temp}). The total
mass required to fit the MIPS data is 1.5$\times$10$^{-3}$
$M_{\earth}$, $\sim$1.9 times lower than the silicate grains due to
the reduced grain density in amorphous carbon. However, we rule out
the possibility that the disk contains solely amorphous carbon as in
that case the emission at 850 \um would extend well beyond the
observed SCUBA emission.

\begin{figure}
\figurenum{15}
\label{radprof_powersize_sil_amc}
\plotone{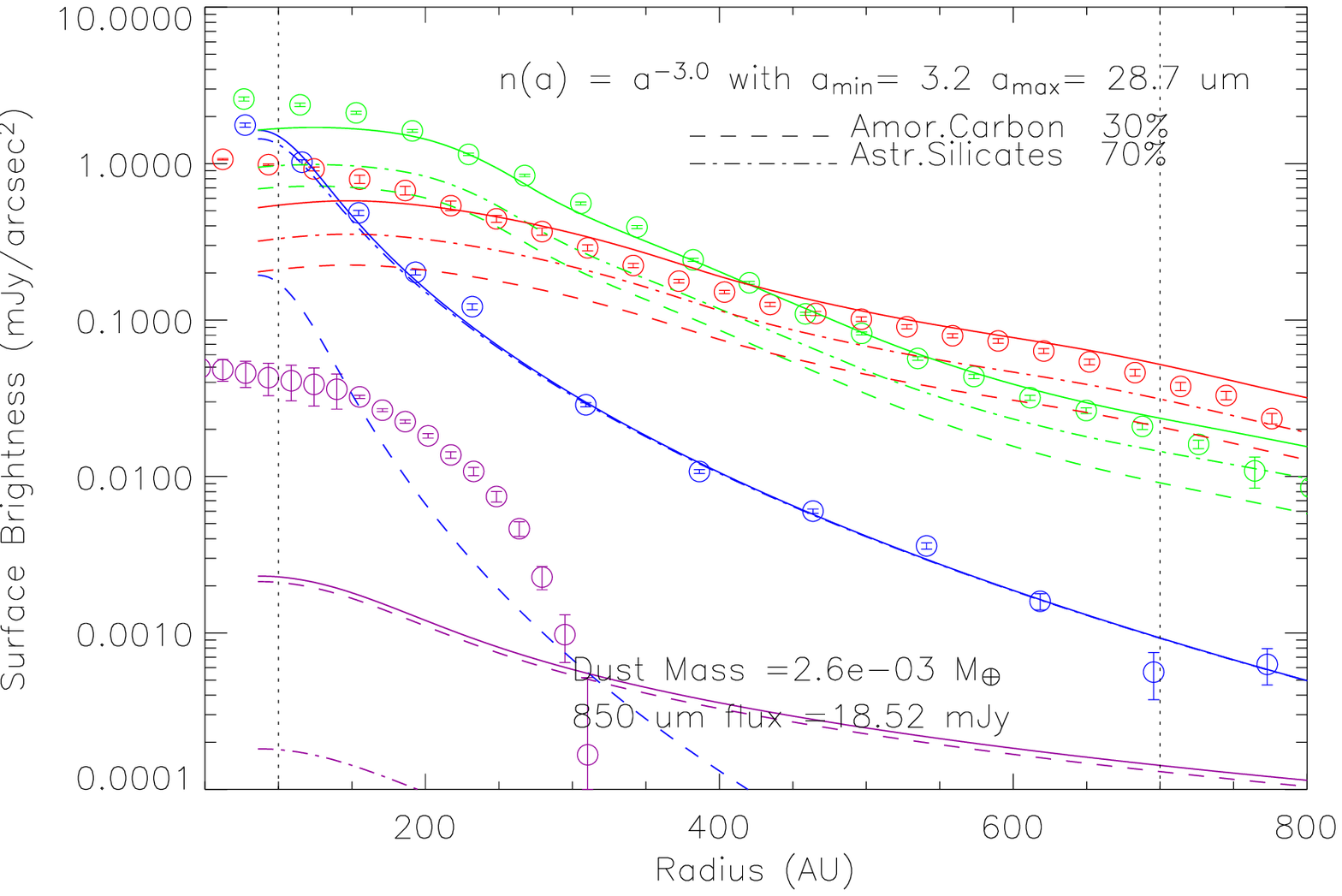}
\caption{Same as Figures \ref{radprof_powersize} and
  \ref{radprof_powersize_amc}, but using a mixture of amorphous carbon
  (30\%) and silicate grains (70\%). The emission from amorphous
  carbon grains is plotted as dashed lines while the emission from
  silicate grains is plotted as dashed-dotted lines. The best fit
  grain size distribution is $n(a) \propto a^{-3.0}$ with a minimum
  size cut-off $a_{min}\sim$3.2 \um and maximum size cut-off $a_{max}\sim$29
  \mm. } 
\end{figure}

A mixture of silicate and carbon grains is probably a more realistic
dust grain model since both silicate and carbonaceous materials have
been found in the Solar System. Without spectral features to identify
grain composition, we simply assume that the particles in the disk
consist of 70\% silicates and 30\% amorphous carbon grains. The best
fit of this admixture model is displayed in Figure
\ref{radprof_powersize_sil_amc} with a grain size distribution $n(a)
\propto a^{-3.0\pm0.6}$ and a minimum size cut-off 3.2$\pm$0.8 \um and
a maximum size cut-off 29$\pm$14 \mm. The derived total mass in the
disk is 2.6$\pm 1.5\times$10$^{-3} M_{\earth}$ with $\sim$18\% from
the amorphous carbon grains. The total emission at 850 \um is $\sim$19
mJy assuming an aperture of 90\arcsec.

The independence of total mass on grain size distribution can be
understood as a consequence of the drop in emission efficiency in
proportion to the grain radius for grains much smaller than the
wavelength; i.e., $Q_{abs} \propto a$ when $a \ll
\frac{\lambda}{2\pi}$ (Rayleigh limit). As a result of this effect,
the radiation goes as surface area times radius, or as $a^{3}$. The
number of grains required for a given surface brightness is therefore
proportional to $a^{-3}$. The mass of a grain goes as volume, or
$a^3$. Therefore, the product of the required number of grains and the
mass per grain is roughly independent of grain size. For grains with
$a \ll \lambda$, the total mass of grains required to produce a given
radiometric signature is roughly independent of $a$. Hence, the system
mass derived from the MIPS data is insensitive to the specifics of the
grain model. Since the temperature distributions for silicate and
amorphous carbon grains are similar, hereafter we assume a pure
silicate composition for our disk model, for simplicity.

\begin{figure} \figurenum{16}
\label{radprof_ring}
\plotone{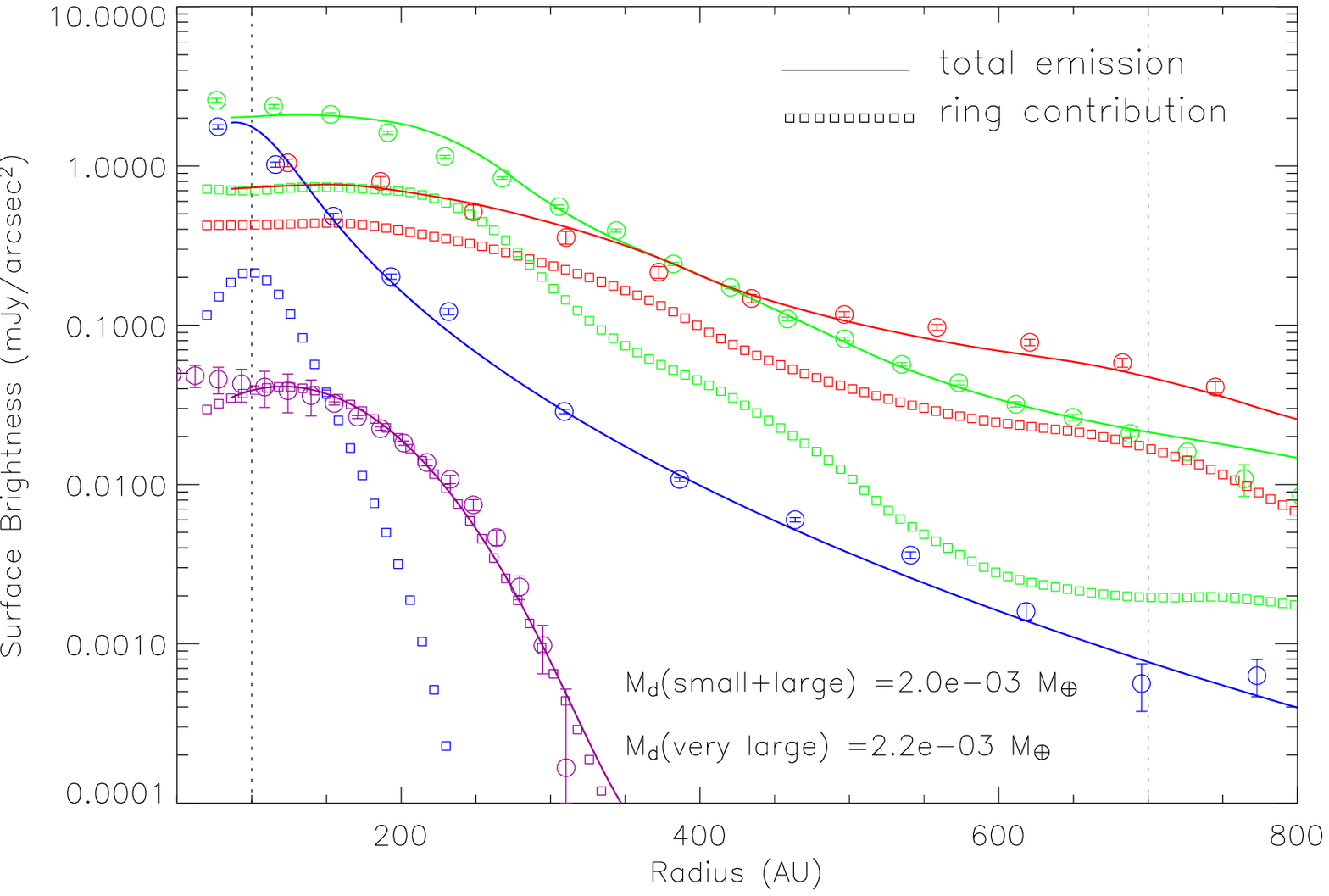}
\caption{Three component model fit with small (2 \mm) and large (18
  \mm) grains in the disk, and very large grain (215 \mm) in the
  ring. The symbols and lines are the same as in Figure
  \ref{radprof_3bands} with colors indicating wavelengths: 24 \mm:
  blue, 70 \mm: green, 160 \mm: red, and 850 \mm: purple. The model
  emission from the 215 \um grain in the ring is plotted as
  dotted-open-squares with color representing different wavelengths.
  The solid lines are the total model emission from the
  three-component model.}
\end{figure}

Our modeled flux at 850 \um is $\sim$4-45 times lower than what was
found in the SCUBA data. This implies that a third component is needed
to account for the 850 \um emission. We believe MIPS is seeing small
grains blown out by radiation pressure while SCUBA is seeing larger
grains that are gravitationally bound to the star and could be the
grinding source of the small grains. The MIPS and SCUBA observations
are then distinctly different grain populations. To further test this
scenario, we assume a ring structure in the disk, containing silicate
grains with $a=$ 215 \mm. We do not try to model the ring's physical
structure; instead we only try to find a structure where its resultant
emission profile is consistent with the 850 \um profile and does not
violate the 24, 70 and 160 \um profiles. We achieved a fit with a ring
structure that has a constant density from 86 to 100 AU, and drops as
$r^{-2}$ from 100 AU to a cut-off radius of 200 AU. The radial
profiles for this three-component model are shown in Figure
\ref{radprof_ring}. In this model, the 24 \um profile is dominated by
the small (2 \mm) grain population in the disk. The 70 \um profile,
while dominated by emission from the large (18 \mm) grains in the
disk, has a significant contribution ($<$200 AU from the star) from
the very large (215 \mm) grains in the ring. Within this radius, the
215 \um grains contribute $\sim$30\% of the total emission.  At 160
\mm, the large grains in the disk and the very large grains in the
ring contribute roughly equal amounts of emission.

Having single-size grains in the ring is not mandatory. For silicate
grains, we found that the modeled 24, 70 and 160 \um radial profiles
remain unchanged for grains with size larger than 180 \um since
$Q_{abs}\sim$1 for very large grains and their corresponding
temperature structure given the same heating source is similar (the
canonical temperature distribution). Therefore, the total mass in the
ring only changes by a factor of $a^1$ since the emission goes as the
grain surface area ($a^{2}$). Due to the uncertainty in the absorption
coefficient in the submillimeter regime
\citep{hildebrand83,pollack94}, we do not try to constrain the
particle size distribution in the ring. The mass in the ring is at
least a few tenths of a lunar mass (on order of $10^{-3} M_{\earth}$),
consistent with the previous submillimeter measurements
\citep{zuckerman93, holland98}.

\subsection{Stochastic Grain Heating}

The relatively constant {\it [24]$-$[70]} color temperature with
radius suggests an alternative class of model in which tiny grains are
heated stochastically by absorbing a single photon from the star.
Stochastic heating occurs when the assumption that the temperature of
a grain is determined by time-averages of the absorption and emission
rates breaks down.  This can occur when the the time interval between the
absorption of photons with energies comparable to or larger than the grain heat
capacity is much longer than the grain cooling timescale, a condition
that can be realized in very small grains ($\lesssim$ a few hundred
\AA). To test the hypothesis of stochastic heating in the Vega system,
we have computed the complete time evolution of grain temperature 
(eg., \citealt{krugel03}) for small grains exposed to the radiation
environment of Vega.  We have chosen a distance of 600~AU,
characteristic of the debris disk extent, for the calculation.

For grains less than about $\sim$50~\AA, the cooling timescale was
computed to be of order a few hundred seconds.  For grains
between 30-50~\AA, the absorption time scale varies from a few tens of
seconds down to a fraction of a second in Vega's radiation field.
With the heating time scale much less than the cooling timescale, the
temperature distributions are very strongly peaked near the
equilibrium temperature and the emission from such grains can be very
well described as an equilibrium process at a single temperature. Of
course, any grain larger than 50~\AA\ will also satisfy the
assumptions of equilibrium heating. At 20~\AA, the heating timescale
has lengthened so that larger temperature fluctuations are observed
and the temperature distribution of the grain has a significant hot
tail.  However, the emission from the grain is still very similar to
that expected from an equilibrium process.  In order to observe
significant radiometric effects of stochastic heating in the Vega
environment, grains of sizes $\le$15~\AA\ are required. In this small
regime, the heating timescales are of order $\ge$ 10$^2$ seconds
and the grain will be stochastically heated.
 
A large number $\sim$10~\AA\ grains (which absorb inefficiently in the
optical and near UV) would be required to absorb sufficient energy
from Vega to account for the infrared excess.  As an illustration, we
carry out a rough order-of-magnitude estimate. We assume that the
grain absorption efficiency for a given wavelength of light goes as
the geometric cross section times the ratio of grain radius to
wavelength. Then, typical stochastically heated 10~\AA\ grains will
have absorption efficiencies well below 1\% of the geometric cross
sections (this estimate is confirmed with real grain optical
properties). If we put the grains at a typical distance of 600 AU, and
assume a density of 3 g/cm$^3$, then we find that the mass of very
small grains required to absorb the fractional luminosity of the star
that is re-emitted in the infrared is within an order of magnitude of
the mass required in grains in our preferred fit using large
grains. It is not clear how to create the necessary numbers of such
tiny grains without producing substantial numbers of larger ones that
would have a strong effect on the radiometric properties of the
system. In addition, \citet{artymowicz88} has calculated the effects
of radiation pressure on such small grains around an A star and finds,
because their absorption cross sections for stellar photons are very
small, that they may not be ejected from the system by radiation
pressure. Thus, the radial extent of the Vega system might be
difficult to explain with tiny grains. The upper limit to the
polarization of the debris system obtained by \citet{mauron98} also argues
strongly against a system dominated by very small grains.  It
therefore is unlikely that a disk dominated by a population of small
stochastically heated grains can explain all of the available Vega
observations; nor is it necessary to invoke such a population as our
preferred model fit using large grains is fully consistent with the
observations.

\subsection{Summary of the Modeling}

A range of models using amorphous silicate and/or carbon grains of
sizes from $\sim$1 to $\sim$50 \um can fit the infrared radiometric
behavior of the disk out to $\sim$800 AU. The exact minimum and
maximum grain size limits depend on the adopted grain
composition. However, all these models require a $r^{-1}$ surface
number density and a total mass of 3.0$\pm 1.5\times$10$^{-3}
M_{\earth}$ in grains. Models that also fit the submillimeter data
additionally require an inner ring (near $\sim$100 AU) with largr
grains ($>$180 \mm) and a total mass $\ge$10$^{-3} M_{\earth}$.

\section{The Origin of the Debris} 
\label{vega_debris_origin} 

One might propose that the dust we see at 100-800 AU is produced {\it
in situ} from a highly extended Kuiper Belt around Vega. Given that
Vega is 2.5 times as massive as the Sun, it is plausible that it could
have a more extended population of such objects than does the solar
system. Following simple mass scaling and assuming a similar surface
density of large bodies in the disks, Vega's Kuiper belt would extend
only about 1.6 times farther than the Sun's (45-55 AU). If Vega's
Kuiper belt extended to 500 AU, its mass would be over 100 times that
of the Sun's Kuiper Belt ($\sim$100$ M_{\earth}$). This mass, and the
implied extent, seem implausible.  Further, the distribution of dust
that would result from {\it in situ} production in such an extended
Kuiper Belt almost certainly would not replicate the smooth surface
brightness distribution we observe. We therefore disregard the
hypothesis that the debris disk around Vega originates from an
extended Kuiper Belt.

The radial symmetry of the Vega disk strongly suggests the star system
itself as the source of the dust. If the dust were of interstellar
origin, it seems likely that it would be inhomogeneously distributed
around the star. Also, \citet{artymowicz97} show that the avoidance
distance around Vega (the distance at which small grains would be
repelled by radiation pressure), is $\sim$3600 AU; therefore, small
interstellar grains would not be present where we see the dust, inside
about 1000 AU.  Further, the radial density distribution of the dust
is consistent with that expected from dust being ejected by radiation
pressure. The debris we detect with MIPS is unlikely of interstellar
origin.

The large extent and the radial and azimuthal distributions of Vega's
disk are all consistent with a model in which the dust originates at a
distance $\sim$100 AU from Vega, and is blown out to larger radii by
radiation pressure.  A further consequence of this model, as will be
shown below, is that the dust we are seeing in the {\it Spitzer} data
originated in an event that took place in the relatively recent past.
These conclusions do not apply to the larger particles in the ring
structure around Vega, which are seen in the submillimeter but have
much less contribution to the flux in the MIPS bands.

The grain-sizes we deduce from our model fits to the data, ranging
from $\sim$1 to $\sim$50 \mm, are consistent with their being pushed
out of the system by radiation pressure, if we assume the larger
grains have significant porosity or are highly non-spherical. The
ratio of the radiation force to the gravitational force on a grain is
parameterized by $\beta = \frac{3L_{\ast} <Q_{pr}>}{16\pi G M_{\ast} c
a \rho}$, where $L_{\ast}$ is the stellar luminosity, $<$$Q_{pr}$$>$
is the radiation pressure efficiency of the grain averaged over the
stellar spectrum ($<$$Q_{pr}$$>\sim$1 when $a \gg $1 \mm,
\citealt{artymowicz97}), $G$ is the gravitational constant, $M_{\ast}$
is the stellar mass, $c$ is speed of light \citep{burns79}. For $\beta
>0.5$, radiation pressure will drive the particle out of the system;
for $\beta < 0.5$, the particle will spiral into the star due to
Poynting-Robertson (P-R) drag. We adopt $\rho$ = 3.5 g/cm$^3$ for
solid ``astronomical silicate'' grains.  Using $L_\ast=60 L_{\sun}$
and $M_{\ast}=2.5M_{\sun}$ for Vega, we find $\beta_{2\mu m}$ = 1.95
and $\beta_{18\mu m}=$ 0.22. The ``small'' grains in our model will be
ejected, while the ``large'' grains will spiral in under the influence
of P-R drag. In order to explain the presence of the larger ($\simeq$
20 \mm) silicate grains at large distances from Vega, we require their
density to be reduced significantly. They may be aggregates of smaller
grains, have significant porosity, or be highly non-spherical. It is
also possible that the smaller ($\simeq$ 2\mm) grains are under-dense,
but that is not required to explain their presence at such large
distances.

Aggregate grains are seen in our own solar system (e.g., cometary dust
and zodiacal dust, \citealt{grun01}), and are a natural consequence of
growth through coagulation in the early Solar Nebula
\citep{kimura02}. Radiation pressure depends not only on the size and
material of the dust grains, but also on their shapes and
structures. Porosity certainly can decrease the bulk density $\rho$,
therefore increasing the $\beta$ value. In order for $a_{eff}$ = 18
\um grains to have $\beta > 0.5$, their porosity has to be greater
than 0.56 (or $\rho_{eff} \sim 1.54$ g/$cm^3$, which is not
unreasonable). If our large grains are very non-spherical (plates,
needles), they could also be ejected by radiation pressure (see
\citealt{il98}, for example).

Particles escaping under the influence of radiation pressure are lost
from the system on timescales of order the orbital period in the
source region \citep{krivov00}. Considerations of energy conservation
lead to the conclusion that such particles reach a terminal radial
velocity given by $v_r \simeq \sqrt{ \frac{2 G M_{\ast}}{r_{initial}}
[\beta - \frac{1}{2}]}$, where $r_{initial}$ is the distance where the
grain is released or produced \citep{amaya05}. This implies velocities
for the escaping particles (assuming $\beta=$1) in Vega's disk in
excess of 5 km/s, if the source region is around 100 AU, consistent
with the location of the submillimeter dust ring. The residence time
for the particles is roughly $t_{residence} \simeq R_{disk}/v_r \leq
1000$ years for a disk radius of 1000 AU. Clearly, if the dust we are
seeing in the {\it Spitzer} images is being blown out by radiation
pressure, the individual dust grains are of recent origin.

For our hypothetical very large grains ($a>$180 \mm) in the inner ring
at 86-200 AU that are seen in the submillimeter data, we find that
$\beta
\sim$0.02, implying very long residence times in that region. They therefore
could be the source for the dust we see at much larger distances, or
could be associated with a population of larger asteroidal bodies
analogous to our own Kuiper Belt, which could be the ultimate source
of the dust at large distances. Cascades of collisions starting with
encounters between these larger bodies in this ring could produce the
small grains that are blown out by radiation pressure force.

The short residence time for small grains ($a \le$50 \mm), combined
with the dust mass we derived in section 5 (see Figures
\ref{radprof_powersize}-\ref{radprof_ring}), places a lower limit on
the dust production rate in the Vega system. Taking the dust mass in
small grains to be $3.0\times 10^{-3}$ $M_{\earth}$ and a residence
time of 1000 years, the required dust production rate is
$\sim$6$\times$10$^{14}$ g/s. \citet{lisse99} estimate an overall dust
production rate of 2$\times 10^8$ g/s for the Comet Hale-Bopp. If the
dust is generated by cometary activity, the Vega debris disk requires
3 million Hale-Bopp-like comets to account for the dust production
rate, which is impossibly high. If we further assume that dust loss
and production have been in approximate balance over the life of the
Vega system (350 My), we find that the mass of the dust lost over that
time is $\simeq 1100$ $M_{\earth}=$ 3.3 $M_{J}$.  This is a lower
limit as we have assumed the longest plausible residence time, and
limited the calculation to only those small grains we are sensitive
to; presumably there are other grains lost to P-R drag in the inner
part of the disk. Further, to produce 3.3 $M_{J}$ worth of dust, a
much larger mass of parent bodies would have to be present in the
disk. Assuming a power law size distribution proportional to
$a^{-3.5}$, where $a$ is the radius (see \citealt{dohnanyi69}), the
mass of the parent bodies would be two to three orders of magnitude
greater, implying a total mass of 330-3300 $M_{J}$ (or 0.3-3
$M_{\sun}$) for Vega's debris disk.

It seems very implausible to have an initial inventory of 330-3300
$M_{J}$ of dust when Vega was born. Only a fraction of the initial
disk mass is presumably incorporated into asteroids; most of it could
have been sequestered in giant planets, accreted onto the star, or
ejected from the system. Besides, debris disks do not form as solid
material only; they start out as massive protoplanetary disks, with
100 times as much gas mass as dust.  The disk masses of Herbig Ae/Be
stars (what Vega would appear as, when first formed) are on the order
of a few tenths of a M$_{\sun}$, measured in the millimeter continuum
\citep{hillenbrand92,mannings97}.  It is very unlikely that this high
dust production rate is in a steady state throughout Vega's life.

The above arguments suggest that the disk we imaged with {\it Spitzer}
is ephemeral. Such a disk could be produced as the result of the
disruption of a large asteroidal or cometary body and subsequent
collisional cascade as pieces of the disrupted body collide with other
bodies in the source region.  The dust mass we observe, 3.0$\times
10^{-3} M_{\earth}$, would form an object $\sim$1000 km in radius (the
size of Pluto) if collected together. Such a large object is unlikely
to undergo a disruptive collision ({\it e.g.}  \citealt{kenyon04}). We
hypothesize instead that a more modest sized object was disrupted, and
the ensuing collisional cascade, interacting with the background
population of asteroidal objects, produced the observed dust. Small
bodies were probably the ultimate source of most of the dust mass,
although regolith ejection from larger bodies \citep{kortenkamp98}
could also have contributed significantly. Thus we favor a scenario in
which the dust we observe is young because it is being blown out of
the system by radiation pressure, but was produced over some finite
interval of time, consistent with a collisional cascade subsequent to
a large disruptive collision. This scenario is also consistent with
the results of \citet{rieke05}, who find that episodic collisions, and
not steady-state production, best explain dust seen around a large
sample of A stars.

\section{Conclusion} 
\label{vega_conclusion}

We have presented images of the Vega debris disk at 24, 70 and 160 \um
obtained by MIPS. The disk is well resolved with 1-$\sigma$ detection
radii of 43\arcsec, 70\arcsec, and 105\arcsec~at 24, 70 and 160 \mm,
respectively. The disk appears circular and smooth at all three
wavelengths. No clumpy structure was found. The surface brightness of
the disk follows a simple $r^{-3}$ (inner disk) or $r^{-4}$ (outer
disk) power law at all three wavelengths, implying the disk density
structure is simple and smooth. A region lacking material that emits
efficiently at MIPS wavelengths, was inferred at a radius of
11\arcsec$\pm$2\arcsec~from a radial profile analysis.

We investigated the nature of the Vega disk with model fitting.  We
found that an axially symmetric, face-on disk with $r^{-1}$ surface
number density can explain the observed radial profiles at all three
wavelengths simultaneously. The observed radial profiles can be well
reproduced with a two-component grain model consisting of a mixture of
small ($a=$ 1.3-2.7 \mm) and large ($a=$ 12-30 \mm) silicate grains
with the former size accounting for $\sim$98\% of the particles in
number. The observed radial profile can also be reproduced with a
power-law grain size distribution, $n(a) \sim a^{-3.0\pm0.5}$ with a
minimum size cut-off 0.7-2.3 \um and maximum size cut-off 35-57
\mm. The minimum and maximum grain sizes in the disk are constrained
by the requirement that the model simultaneously reproduce the 24, 70,
and 160 \um radial profiles.

Silicate and carbonaceous grains have very similar radial temperature
profiles in the radiation environment around Vega.  Therefore, while
the size distribution is well constrained by the surface brightness
profiles, the grain composition is not. While the composition of the
dust grains in the disk is unknown, it is fully consistent with a
mixture of silicate and carbonaceous dust. Regardless of the exact
composition, the total mass of the dust seen by MIPS is
$\sim$3.0$\times$10$^{-3}$ $M_{\earth}$ assuming an outer radius of
1000 AU.

The total flux at 850 \um emitted by the grains required to fit the
MIPS data is much less than what was measured in the SCUBA
data. Another dust component is needed to account for the emission at
850 \mm. Emission from a ring composed of grains larger than 180 \um
can account for the observed 850 \um profile, while their contribution
in the MIPS wavelengths does not change the disk model profiles.

The ring-like structure in our disk model represents a reservoir for
gravitationally bound planetesimals, where collisions can occur and
generate small dust grains as debris. The dust production rate implied
by our observations, $\sim$10$^{15}$ g/s, would require this
asteroidal reservoir to be improbably massive, were the disk we
observe with {\it Spitzer} in a steady state. We conclude that the
disk is ephemeral, the consequence of a large and relatively recent
collisional event, and subsequent collisional cascade. Radiation
pressure sets the small dust grains produced in these collisions on
unbound hyperbolic trajectories and they stream out through the disk
and leave the system with a time scale on the order of $\sim$1000
yr. The large extent of the disk as seen by MIPS is consistent with
the dust being ejected by radiation pressure.

\acknowledgments

We thank the MIPS instrument team (especially James Muzerolle, Jane
Morrison and Chad Engelbracht) for assistance with data analysis
and useful comments.  We are grateful to David Frayer and  Deborah
Padgett in the {\it Spitzer} Science Center for support on data
reprocessing. We thank Mike Jura for suggestions and comments.
Support for this work was
provided by NASA through Contract Number 960785 issued by JPL/Caltech.

\begin{deluxetable}{lccccccc}
\tablecaption{Observing Log\label{obslog}}
\tablewidth{0pt}
\tablehead{
\colhead{Array} & \colhead{Field Size} & \colhead{Pixel Size} & \colhead{Exposure}  &
\colhead{\# of Pos.} & \colhead{Eff. Exp.}& \colhead{1-$\sigma$ per pixel}\\
\colhead{}      & \colhead{}   & \colhead{arcsec}         & \colhead{exp x cyc.} & 
\colhead{}      & \colhead{sec} & \colhead{mJy/arcsec$^2$} 
}
\startdata
A24      & small & 2.55 & 3 s x 1      &  4  & 120$\sim$126\tablenotemark{a}  &  0.01 \\
A70F     & small & 5.20 & 3 s x 1      & 12  & 132$\sim$151  &  0.13 \\
A70      & large & 9.98 & 3 s x 8      &  1  & 195$\sim$255  &  0.04 \\
A160     & large & 16$\times$18 &3 s x 4      &  4  & 38$\sim$66  &  0.02 \\
\enddata
\tablenotetext{a}{The typical effect exposure time is $\sim$ 63 sec
  for the central 6\arcsec~region due to the saturation.}
\end{deluxetable}

\end{document}